\documentclass[aps,pre,reprint,superscriptaddress]{revtex4-2}

\usepackage[T1]{fontenc}
\usepackage[utf8]{inputenc}
\usepackage[english]{babel}
\usepackage{amsmath,amssymb,bm}
\usepackage{graphicx}
\usepackage[version=3]{mhchem}
\usepackage{xcolor}
\usepackage{hyperref}

\begin{document}

\title{Granular structure and heterogeneous deformation preceding avalanches}

\author{Ibrahim Awada}
\author{Michel Bornert}
\author{Vincent Langlois}
\author{Julien L\'eopold\`es}
\email{julien.leopoldes@univ-eiffel.fr}
\affiliation{Navier, CNRS, Univ Gustave Eiffel, ENPC, Institut Polytechnique de Paris, Marne-la-Vall\'ee, France}

\begin{abstract}
Granular packings undergo intermittent rearrangements, known as precursors, before avalanche onset. Using grain-scale tracking in an inclined granular layer, we show experimentally that the number and recurrence of precursors change with the compatibility between the anisotropic structure inherited from preparation and the subsequent loading direction. Preparation also modifies the propagation of the mobility front, defined as the region in which the packing undergo measurable displacement: frequent precursors produce small progressive advances, whereas rare precursors involve larger extensions. By contrast, several features remain unchanged across protocols. The ensemble-averaged displacement always decays exponentially with depth over approximately four grain diameters, while the local shear-strain variance follows a common quadratic scaling with the mean strain. A two-state model suggests that, regardless of the packing structure, local strain heterogeneities affect approximately one quarter of the elementary cells, superimposed on the exponential strain profile. The reported results separate preparation-dependent precursor activity from the robust spatial features of the deformation preceding avalanches.
\end{abstract}

\maketitle

\section{Introduction}
Solids begin to flow when the applied stress reaches the static yield threshold. This solid-to-liquid transition is directly relevant to numerous industrial sectors, ranging from food science and civil engineering to the pharmaceutical industry, as well as to geophysical processes. At the macroscopic scale, strain softening or strain-rate softening may trigger a mechanical instability at the threshold. This instability manifests itself through a heterogeneous strain field and eventually leads to strain localization.

However, even below the macroscopic yield threshold, plastic deformation is already spatially heterogeneous~\cite{Kuhn1999}. As the applied stress approaches the threshold, the characteristic length scale associated with these heterogeneities, such as 
the correlation length, increases strongly and may diverge with specific critical exponents. This behavior has received considerable attention because the transition between solid-like and flowing states bears some of the signatures of a dynamical phase transition~\cite{nicolas2018deformation}. Such a framework has been proposed to describe the response of a broad range of disordered systems, including ice~\cite{weiss2003threedimensional}, foams~\cite{gopal1995nonlinear}, colloidal suspensions~\cite{weeks}, gels~\cite{buisson2003intermittency}, frictional interfaces~\cite{carobaumb}, and granular materials~\cite{vilotte}.

Understanding these precursory deformations is therefore of fundamental interest since they provide information about the collective mechanisms through which a mechanically stable solid loses stability. They are also relevant to hazard assessment, since many geophysical failures, including earthquakes and landslides, involve the progressive destabilization of frictional and granular materials~\cite{scholz2019}. For instance, field observations of unstable slopes reveal an acceleration of deformation before macroscopic failure~\cite{petley2002} or an increase in the proportion of large events of acoustic emission~\cite{amitrano2005}. However, the interpretation of these measurements beyond monitoring remains difficult because of the complexity of natural slopes.

Laboratory experiments and numerical simulations provide controlled model systems in which the approach to failure can be studied with spatial and temporal resolution. In sheared granular layers, Nasuno \textit{et al.} observed intermittent microscopic slip events during the loading phases preceding macroscopic sliding~\cite{nasuno1998}. Discrete-element simulations subsequently showed that microslip activity may accelerate as a major slip event is approached~\cite{ferdowsi2013}. In inclined granular packings, numerical simulations have revealed spatially heterogeneous pre-avalanche instabilities~\cite{vilotte}, while experiments have identified recurrent rearrangements, termed precursors, that occur before avalanche onset~\cite{nerone2003}. Full-field measurements further demonstrated that the pre-avalanche deformation is both intermittent and spatially heterogeneous~\cite{amon2013}. Precursors also depends on material and preparation parameters, including friction, vibrations, and packing fraction~\cite{gravish2014,oger2007disorder,kiesgen12}.

Despite these advances, the spatial distribution of strain during individual precursor events remains only partially characterized. Moreover, the role of the internal structure inherited from the preparation of the granular packing is not well established. In this work, we use a preparation protocol that allows us to address these two questions. \textcolor{black}{Using grain-scale tracking, we characterize the number and recurrence of precursors, as well as the associated deformation field and its spatial fluctuations. We examine how these properties depend on the initial anisotropy of nearest-neighbor orientations within the packing.}

\section{Experimental Methods}
The experiment is conducted in a parallelepipedic cell containing an immersed granular medium. Following a series of preparation protocols designed to control the granular structure, the cell is progressively inclined to shear the material until avalanche onset. Grain tracking through the sidewall of the cell is used to characterize both the local structure, through the computation of both first-neighbour polar distributions and shear strain.

\subsection{Preparation of the packings and shear direction}
\begin{figure}[h]
\centering
  \includegraphics[width=9cm]{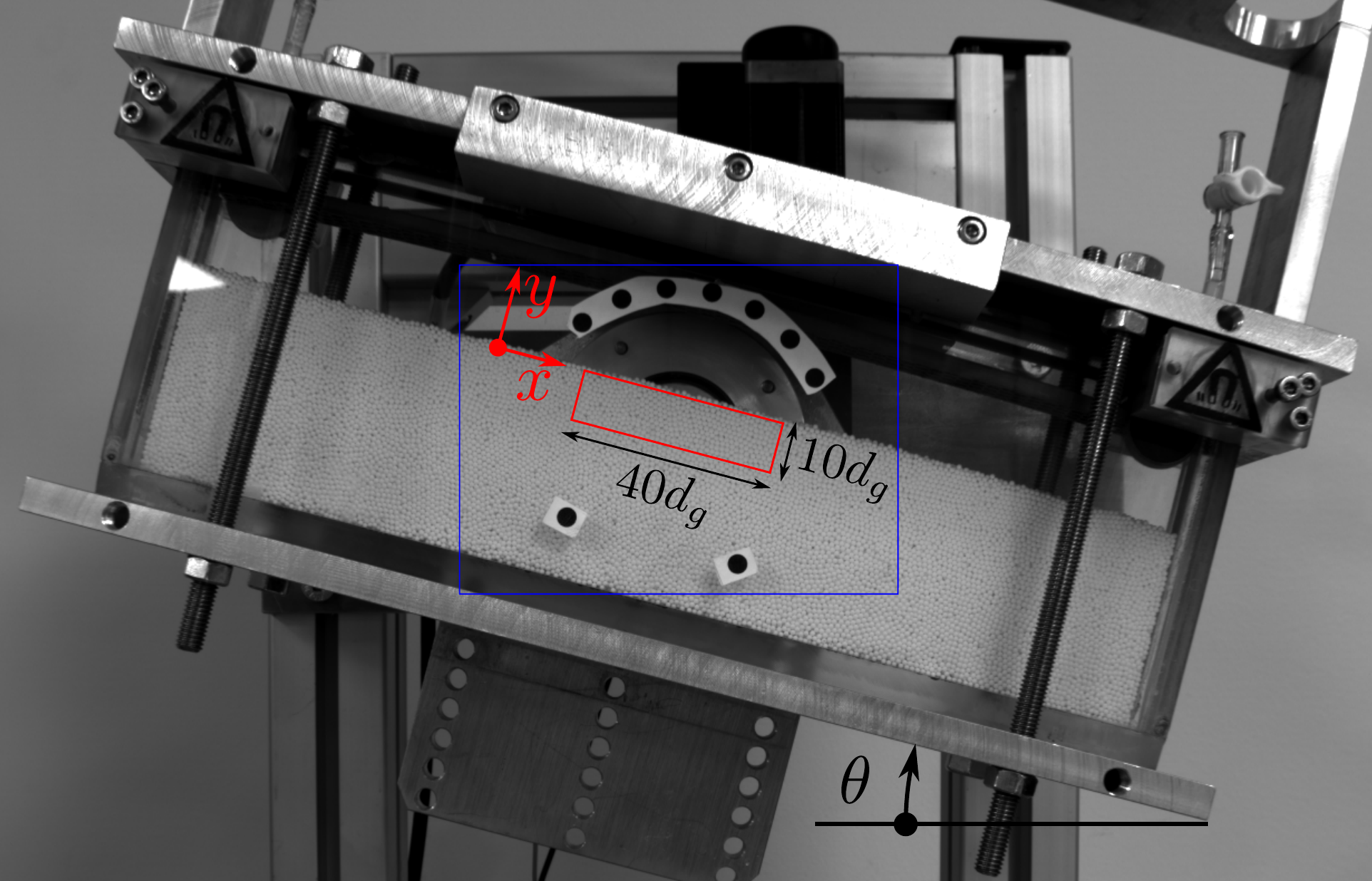}
  \caption{Experimental setup. The cell is filled with the ceramic grains and rotated by an angle $\theta$ with the horizontal. The blue box indicates the camera’s field of view, which spans ($86d_g \times 64d_g$) (see text).
 \textcolor{black}{The red box indicates the region in which the granular structure and local deformation were analyzed.}.
 }
  \label{fig:exp_setup_precurseur}
\end{figure}

We use a transparent rectangular PMMA cell (Fig.~\ref{fig:exp_setup_precurseur}) with dimensions \(30~\mathrm{cm}\) (width), \(10~\mathrm{cm}\) (height), and \(2~\mathrm{cm}\) (thickness). Before being mounted on a rotating platform, the cell is partially filled with monodisperse ceramic beads ($d_g = 1.50 \pm 0.05~\mathrm{mm}$), after which the remaining volume is filled with water. \textcolor{black}{This suppresses both the cohesion caused by capillary bridges in air and ageing effects associated with capillary condensation.} 

\begin{figure}[h]
\centering
  \includegraphics[width=9cm]{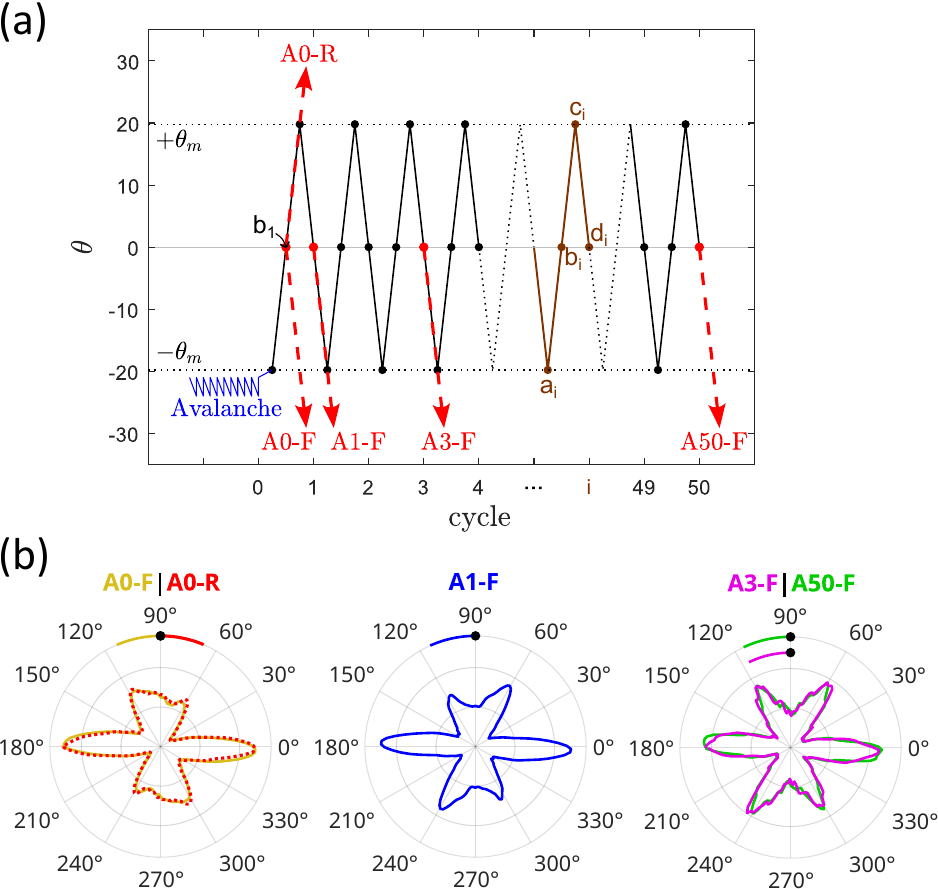}
    \caption{(a) Various steps used to control the structure of the granular. The preparation avalanches (shown in blue) are followed by $N$ oscillations. Then, the samples are inclined (dashed red arrows) until the avalanhe onset either in the opposite direction (``reverse'' loading) or in the same direction (``forward'' loading) as the preparation avalanches. $a_i$, $b_i$, $c_i$ and $d_i$ indicate the positions at which images are taken.
  (b) Polar diagrams showing the angular distribution of first-neighbor positions prior to tilting to the avalanche onset. The loading path is indicated by small circular arcs using the same color as the corresponding polar diagram. A0-F and A0-R are in position $b_1$, AN-F in position $d_\mathrm{N}$.}
  \label{fig:protocol}
  \end{figure}

The preparation of the samples starts with a rapid clockwise rotation of the cell by at least $180^\circ$. This stage is followed by a slow continuous counterclockwise rotation during which the packing undergoes an avalanche at $-\theta_a$ and then relaxes to the angle of repose, $-\theta_r \approx 21.0^\circ$. This continuous rotation erases the memory of the initial configurations and yields a reproducible avalanche angle\cite{evesque1993piledensity}. It induces approximately twenty successive avalanches, hereafter referred to as ``preparation avalanches''. After the preparation avalanches, the samples are subjected to N$=0,1,3~\text{or}~50$ oscillations at $\theta_{m} = \pm19.8^\circ$ (below the angle of repose) as shown in Fig.~\ref{fig:protocol}a. The samples are then brought to the horizontal position labelled $b_1$ for $N=0$ or $d_\mathrm{N}$ otherwise. Finally, they are tilted until the avalanche onset, either in the same direction as the preparation avalanches (forward direction ‘F’) or in the opposite direction (reverse direction ‘R’). We use the following notation to indicate which protocol was used: 

- AN-F. After the preparation avalanches, the cell is subjected to $N$ oscillation cycles and then tilted in the forward direction. We have prepared packings with N$=0,1,3~\text{and}~50$. 

- A0-R. After the preparation avalanches, these samples are not subjected to any oscillation cycle and are tilted in the reverse direction.

To ensure the robustness of the results and evaluate the reproducibility of the observed phenomena, each experiment is repeated at least ten times.

\subsection{Imaging and strain calculation}
\label{notations}
To study grain rearrangements during the experiments, we used a camera (Allied Vision GX3300, 3296 $\times$ 2472 pixels $\equiv 86d_g \times 64d_g$) to image the grains located along the front wall of the cell. The camera was positioned outside the cell, facing the transparent wall. Image acquisition was synchronized with the mechanical control system through a centralized Python interface. For all samples, shear loading was applied by quasi-statically tilting the cell in angular increments of 
($\delta\theta=0.1^\circ$). \textcolor{black}{The prescribed angular acceleration and deceleration were $0.2^\circ\,\mathrm{s}^{-2}$. Over increment, the motion lasted approximately $1.4$~s, with mean and maximum angular velocities of approximately $0.07^\circ\,\mathrm{s}^{-1}$ and $0.14^\circ\,\mathrm{s}^{-1}$, respectively.}  After each increment, the rotation was stopped for 2 s. \textcolor{black}{This pause ensures that any residual motion following a precursor has ceased before the acquisition of 10 optical images and averaging.}

Grain detection was performed using an automated image-analysis protocol based on deep learning and optimized for dense granular media. \textcolor{black}{Only grains in the layer directly adjacent to the transparent wall and having fully visible contours were retained for the two-dimensional	analysis.} The procedure consists of supervised segmentation using a StarDist neural network\cite{stardist}, and determination of grain centers\cite{awada_2025}. As the cell rotates, the entire granular packing appears to translate and rotate in the images. To work in a reference frame attached to the cell, the bead positions were first rotated according to the angular position of the rotating stage. The remaining translation was determined from the average position of grains located far from the free surface, which remained immobile during the experiments. The black dots placed on the cell wall, visible in the photograph of the experiment in Fig.~\ref{fig:exp_setup_precurseur}, were also used to confirm the validity of this correction method. This procedure, together with the grain-tracking accuracy, leads to a typical noise floor in grain displacement of $u_{\text{noise}}\approx0.2~$pix. 

Then, the grain-scale deformations at each loading step were determined from grain displacements between successive images using the micromechanical approach developed by Bagi~\cite{Bagi1996}. Further details can be found in~\cite{Awada2025}. 
As mentioned in the introduction, the precursors of granular avalanches studied in this paper are intermittent and can occur at any inclination angle below $\theta_a$. They are detected following the procedure explained later in~\ref{procedure:detect}. To track precursor activity as a function of the inclination angle, we detect the first and last precursors and compute the corresponding mean occurrence angle. Moreover, some other precursors can be grouped in small equally spaced angular windows between the first and last events. Configuration average, corresponding to average of the displacement or strain at a given location over repeated experiments, is denoted by $\langle \cdots \rangle$. Spatial average is denoted by $\overline{\cdots}$ and, unless specified otherwise, is calculated within a region of interest of size ($40d_g \times 10d_g$) (red rectangle in Fig.~\ref{fig:exp_setup_precurseur}).

\subsection{Characterization of the granular structure}
We have characterized the orientation of the nearest neighbors over the region of interest of size \(40d_g \times 10d_g\) (red rectangle in Fig.~\ref{fig:exp_setup_precurseur}). \textcolor{black}{Two grains, $i$ and $j$, are considered close neighbors if the distance between their centers $d_{ij}$ satisfies the condition $d_{ij} - (d_i + d_j)/2 < \text{1.5 pixel}\approx\sigma_d$, where $d_i$ and $d_j$ are the diameters of the two grains (determined from the StarDist segmentation), and $\sigma_d$ denotes the standard deviation of the grain diameters.}
From the grain-center coordinates, the orientation of each first-neighbor pair was calculated relative to a horizontal axis in a reference frame attached to the cell. The angular range was divided into equal sectors, and the number of neighboring grains falling within each sector was counted. The resulting distributions are presented as polar diagrams (a circular polar diagram indicates isotropy, whereas angular maxima reveals preferential neighbor orientations).

The polar diagrams in Fig.~\ref{fig:protocol}b correspond to the structure of the packings  prior to tilting to avalanche onset. For every protocol, a large proportion of neighboring grains are located parallel to the free surface, as indicated by the dominant lobes near $0^\circ$ and $180^\circ$. This layering effect, previously observed in monodisperse packings under shear flow~\cite{TsaiGollub2004,WegnerEtAl2014}, corresponds to chain-like structures aligned with the streamlines generated by the preparation avalanches. In addition, all polar diagrams exhibit secondary lobes near $60^\circ$, $120^\circ$, $240^\circ$, and $300^\circ$, indicating an overall tendency toward hexagonal ordering, which is promoted by the monodispersity of the grain diameters.

Importantly, Fig.~\ref{fig:protocol}b \textcolor{black}{also reveals differences} between the structures provided by the different protocols. 
\begin{table}[htbp]
\centering
\setlength{\tabcolsep}{4pt}
\begin{tabular*}{0.48\textwidth}
{@{\extracolsep{\fill}}lccccc}
\hline
\textbf{Protocol}
& $\theta_a$ ($^\circ$)
& $n_p$
& $\theta_{\mathrm{first}}$ ($^\circ$)
& $\mu_{ra}$ \\
\hline
A0-R
& $23.1$
& $16.0 \pm 1.8$
& $4.0$
& $0.104$ \\
A0-F
& $23.3$
& $0.7 \pm 0.5$
& $21.9$
& -- \\
A1-F
& $24.4$
& $8.8 \pm 1.0$
& $8.0$
& $0.133$ \\
A3-F
& $25.5$
& $4.6 \pm 1.2$
& $17.7$
& $0.058$ \\
A50-F
& $26.5$
& $2.4 \pm 0.5$
& $23.3$
& $0.039$ \\
\hline
\end{tabular*}
\caption{For each protocol, the table gives the avalanche angle $\theta_a$, the average number of precursors observed during inclination from $0$ to $\theta_a$, and the average angle $\theta_\mathrm{first}$ at which the first precursor is detected. The last column gives the parameter $\mu_{ra}$ obtained from the empirical fit of the surface displacement as a function of the inclination angle, $u_x(0,\theta) \approx u_{\mathrm{first}}\exp\left(\frac{\mu-\mu_{\mathrm{first}}}{\mu_{ra}}\right)$ with $\mu=\tan\theta$ (see text).}
\label{tab:proppacking}
\end{table}
The amplitude of the $120^\circ$ lobe in samples A0-R and A0-F is larger than the amplitude of the $60^\circ$ lobe. This reveals higher density of first neighbours along the  $120^\circ$ direction. Fig.~\ref{fig:protocol}b shows that this anisotropy is modified by a single oscillation at $+\theta_m$ since the $60^\circ$ lobe is now dominant over the $120^\circ$ one for the A1-F samples. Moreover, increasing the number of oscillation cycles provides a more symmetric structure as shown for A3-F and A50-F for which $60^\circ$ and $120^\circ$ are hardly distinguishable. With increasing N, the evolution of the polar diagrams towards a symmetric shape is also accompanied by a progressive compaction\cite{Awada2025}. To support this interpretation, Figure~\ref{fig:freesurf} in the Supplemental Material shows the evolution of the free-surface level with the number of oscillations, which is used as a proxy for changes in the global packing fraction.

\section{Results}
\subsection{Number of precursors}
\label{procedure:detect}
\begin{figure}[htbp]
\centering
  \includegraphics[width=9cm]{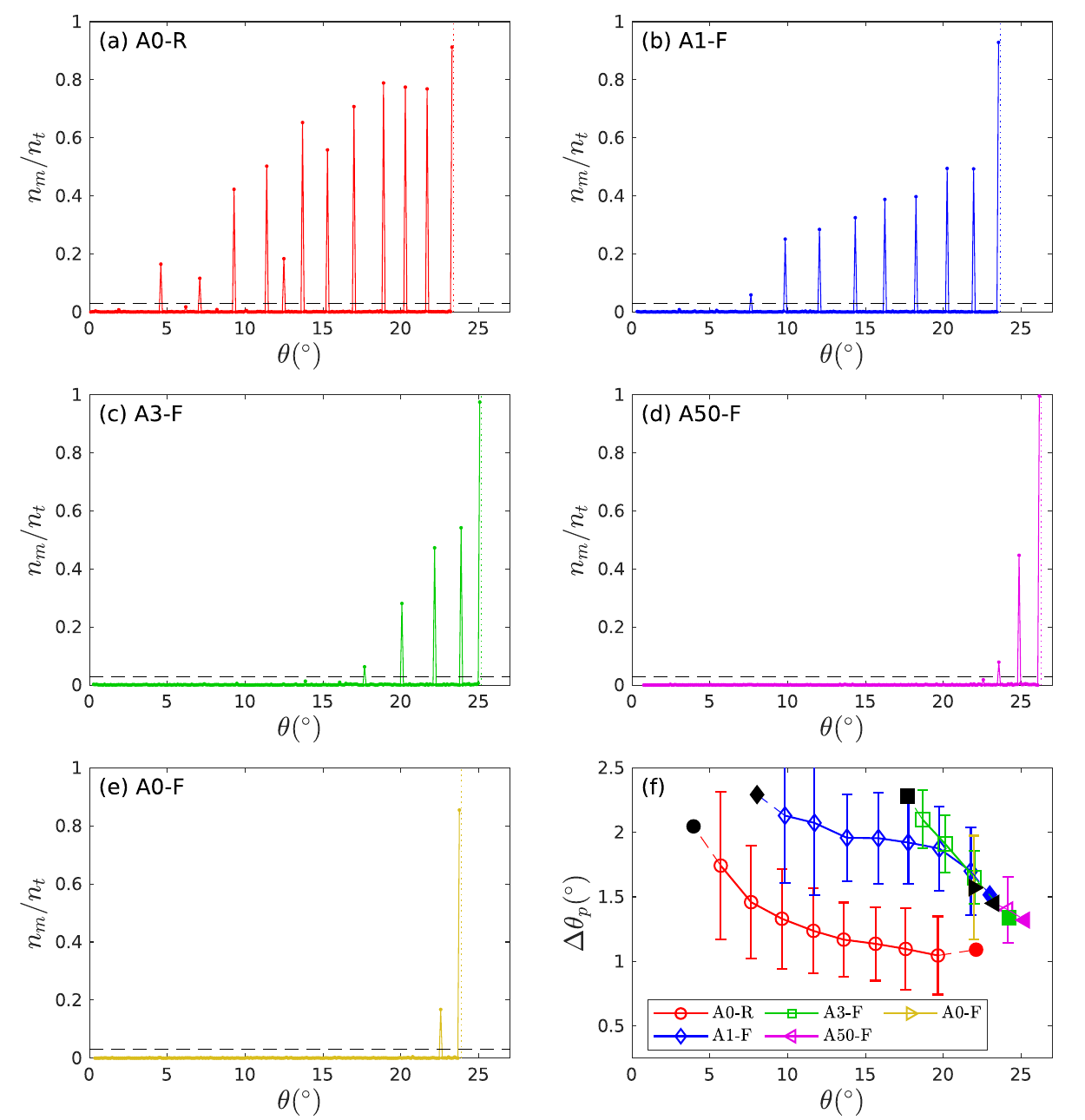}
  \caption{For a single experiment, number of mobile grains $n_m$, normalized by the total number of grains $n_t$, as a function of the inclination angle. (a), (b), (c), (d) and (e) correspond to A0-R, A1-F, A3-F, A50-F and A0-F respectively. For each experiment, the last high peak is associated to the avalanche. The horizontal dashed line, located at $n_m/m_t=3\%$, represents the detection threshold for precursors. (f) Additional inclination required to trigger the next precursor as a function of the inclination angle. The black symbols indicate the first precursor, while the filled colored symbols indicate the last precursor.
}
  \label{fig:pics}
\end{figure}
\textcolor{black}{The precursors, defined as macroscopic events involving a large part of the system, were detected in a large region of size $80d_g\times15d_g$.} For each inclination angle, the displacement vector $\mathbf{u} = (u_x, u_y)$ of all grains in this region is computed along with the intensity $\|\mathbf{u}\| = \sqrt{u_x^2 + u_y^2}$. The grains are classified as mobile if $\|\mathbf{u}\| > 3 \times u_{\text{noise}}$. The proportion of mobile grains, defined as the ratio between the number of mobile grains $n_m$, to the total number of grains $n_t$ \textcolor{black}{located within the large box $80d_g\times15d_g$,} is plotted in Fig.~\ref{fig:pics} for selected experiments, one for each preparation protocol. \textcolor{black}{In the following, we focus on events for which the proportion of mobile grains exceeds $3\%$, hereafter referred to as precursors.
}
All graphs exhibit well-defined peaks, that have been called avalanche precursors in previous studies~\cite{nerone2000}. First, the number of oscillation cycles, $N$, has a significant effect on the number of precursors, as evidenced by the decrease in the number of precursors for protocols A$N$-F with $N = 1, 3, 50$ (Fig.~\ref{fig:pics}b, c and d respectively). Accordingly, the average number of precursors $n_p$ detected before the avalanche decreases from approximately $8.8$ for A1-F to $2.4$ for A50-F (Table~\ref{tab:proppacking}). Note that A0-F is characterized by less than one precursor per experiment (see also Fig.~\ref{fig:pics}e), although its structure before shear loading is identical to A0-R, which is the protocol producing the largest number of precursors ($16.0$). 

As expected, Table~\ref{tab:proppacking} shows that a smaller number of precursors during the tilting phase preceding the avalanche is associated with a larger onset angle $\theta_{\mathrm{first}}$. This analysis can be complemented by considering the angular interval $\Delta\theta_p$ between successive precursors~\cite{nerone}, shown in Fig.~\ref{fig:pics}f. For protocols A0-R, A1-F, and A3-F, for which $\theta_{\mathrm{first}} \ll \theta_a$, one finds $\Delta\theta_p \approx 2^\circ$. As the avalanche angle $\theta_a$ is approached, $\Delta\theta_p$ decreases to approximately $1^\circ$ for all samples. Protocol A0-R, which exhibits the largest number of precursors and the smallest value of $\theta_{\mathrm{first}}$, also displays the shortest angular intervals, corresponding to the most frequent precursor activity.

The distinct precursor activities observed for the different protocols can be related to specific features of the displacement and strain fields, as explained in the next paragraphs.

\subsection{Displacement profiles}
\label{section:av_grain_motion}
\begin{figure}[htbp]
\centering
  \includegraphics[width=6cm]{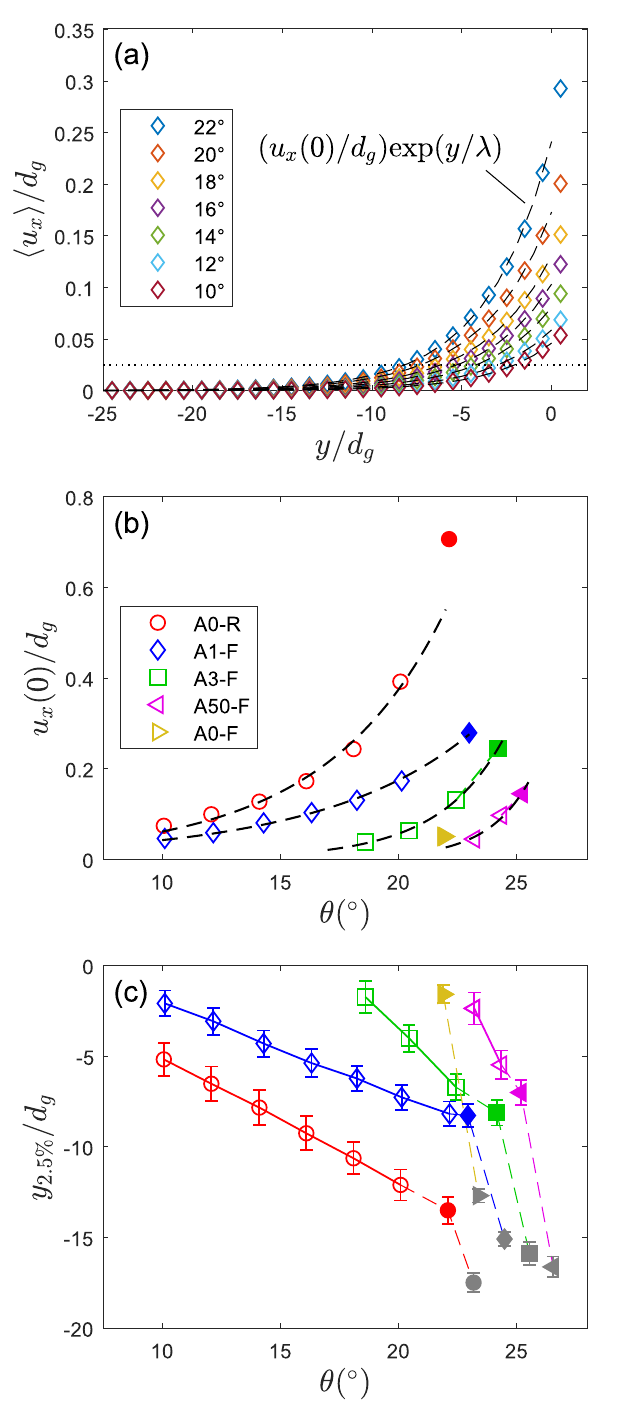}
  \caption{(a) Displacement profile for protocol A1-F. The mean displacement,
$\langle u_x \rangle$, at depth $y$ below the free surface is normalized
by the grain diameter, $d_g$. (b) Transverse displacement at the free surface,
$u_x(0)$, for all samples as a function of the inclination
angle. The filled colored symbols indicate the last precursor. \textcolor{black}{The dashed curves are obtained by fitting the expression $u_n \exp\left(\left(\tan\theta-\mu_0\right)/\mu_{ra}\right)$ to the experimental data.} (c) Depth at which the mean grain
displacement exceeds $2.5\%$ of the grain diameter, $y_{2.5\%}$, plotted as a function
of the inclination angle. Same color code as in (b). The filled colored symbols indicate the last precursor. The filled grey symbols are associated to the avalanches.
The error bars represent the difference between the values of $y_{2.0\%}$ and $y_{3.0\%}$.}
\label{fig:profile}
\end{figure}
\textcolor{black}{The displacement profiles were computed over a window extending to a depth of $25d_g$, in order to capture their decay and determine their functional dependence on depth.} During the precursor events, the transverse displacement of the grains parallel to the free surface, $u_x$, is at least ten times greater than the normal displacement, $u_y$. \textcolor{black}{We focus on transverse displacement and Fig.~\ref{fig:profile}a shows that it} decreases exponentially with depth such that $\langle u_x(y,\theta)\rangle/d_g\approx \left(u_x(0,\theta)/d_g\right)\exp(y/\lambda)$ where $\lambda$ is the cutoff length. This exponential profile is observed for all angles of inclinations, and all protocols with a consistent $\lambda \approx (4\pm0.5)d_g$. \textcolor{black}{The displacements primarily result from frictional sliding and grain rearrangements, which produce grain-center displacements without significant deformation of the grains themselves because of the high Young's modulus $\simeq 400$~GPa\cite{Munro1997} of the grains.}

Moreover, the surface displacement increases with the inclination angle (Fig.~\ref{fig:profile}b) and we find that it can be described by $u_x(0,\theta) \approx u_{\mathrm{first}}\exp\left(\frac{\mu-\mu_{\mathrm{first}}}{\mu_{ra}}\right)$ with $\mu=\tan\theta$. Here, $u_{\mathrm{first}}$ is 
the \textcolor{black}{typical displacement of the smallest precursors detected, $\mu_{\mathrm{first}}=\tan \theta_{\mathrm{first}}$ is the threshold below which no precursor occurs}, and $\mu_{ra}$ is a rate parameter that describes how fast the surface displacement increases once the threshold is exceeded. Low $\mu_{ra}$ implies a rapid increase of surface displacement. 

The expression for $u_x(0,\theta)$ therefore contains two parameters. The value of $u_{\mathrm{first}}$ varies only slightly between protocols, yielding $u_{\mathrm{first}}/d_g \approx 0.03 \pm 0.01$. The fitted values of $\mu_{ra}$ are reported in the last column of Table~\ref{tab:proppacking}. Protocols such as A3-F or A50-F exhibiting ``late'' precursors, characterized by large values of $\theta_{\mathrm{first}}$, show a rapid increase in surface displacement with inclination, as reflected by their low values of $\mu_{ra}$. By contrast, the surface displacement increases \textcolor{black}{more} gradually for A0-R and A1-F, which have smaller values of $\theta_{\mathrm{first}}$, indicating a more progressive approach to avalanche onset.
\subsection{Strain maps}
\begin{figure}[htbp]
 \centering
 \includegraphics[height=14cm]{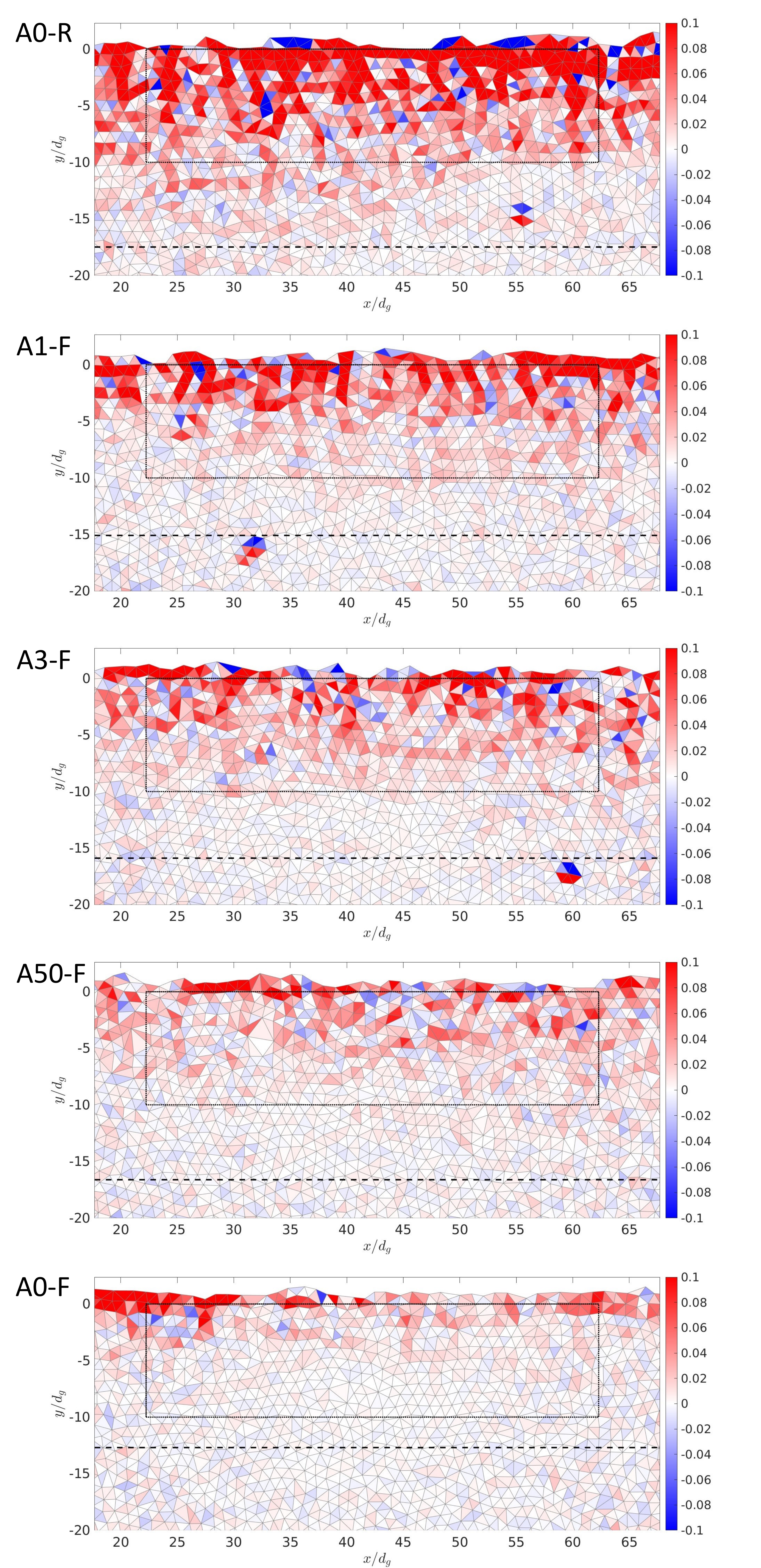}
 \caption{Strain maps at the last precursor event before avalanche. From from top to bottom : A0-R, A1-F, A3-F, A50-F and A0-F. The horizontal dashed black lines correspond to the mean depth at which deformation localizes during an avalanche quantified by $y_{2.5\%}$. The black box delineates the analysis region used to compute the displacement and strain profiles. In order to facilitate the comparison of shear maps, we have reversed the signs of the shear in the case of the A0-R protocol. In these plots, the Bagi-cell strain scale shown by the color bar ranges from $-0.1$ to $+0.1$.
}
 \label{fig:strainfield}
\end{figure}

Fig.~\ref{fig:strainfield} shows, for each protocol, the shear strain fields associated with the last precursor preceding the granular avalanche, i.e. when the pile lies just below the solid-liquid transition. Regardless of the protocol, the strain field is heterogeneous at the scale of the grains, with patches of Bagi cells exhibiting intense shear ($\epsilon_{xy}\approx -0.05$) surrounded by regions of weaker, or even positive, strain (note that such an heterogeneity is also observed at lower inclination angles). At the same time, the strain patterns differ markedly from one protocol to another. Indeed, in A0-R, patches of large deformation extend down to approximately $\sim 13d$, whereas in A50-F, intense deformation remains confined to the vicinity of the free surface. 

The strain maps suggest that deformation vanishes beyond a well defined plane. The depth at which this occurs can be tracked as a function of inclination. To this end, the analyzed region is divided into horizontal slices of height $d_g$, and the deformation is averaged spatially within each slice and across all samples prepared with the same protocol. The depth $y_{2.5\%}$ at which the mean displacement exceeds $2.5\%\,d_g$ is taken as the position of a mobility front separating the deformed and undeformed regions, and is plotted in Fig.~\ref{fig:profile}c as a function of inclination.

For each preparation protocol, $y_{2.5\%}$ decreases linearly with increasing inclination, indicating the propagation of a mobility front as already seen in previous studies~\cite{amon2013,gravish2014}. This front propagates at a rate that depends on the preparation protocol. Protocols with many precursors such as A0-R and A1-F have slow mobility fronts while protocols with few precursors have fast mobility fronts. However, as shown by the grey points in Fig.~\ref{fig:profile}c, the thickness of the sheared zone during avalanche flow $y_a=|y_{2.5\%}|\approx 15d_g$ do not depend significantly on the preparation protocol. 
For comparison, this mean avalanche-front position is also indicated by the dashed black line in Fig.~\ref{fig:strainfield}. Therefore, in our experiment, we find that strain heterogeneities preceding the avalanche and strain localization at avalanche are spatially decoupled, probably governed by distinct mechanisms\cite{kabla05}.
\subsection{Strain fluctuations}
\begin{figure}[htbp]
 \centering
 \includegraphics[height=5cm]{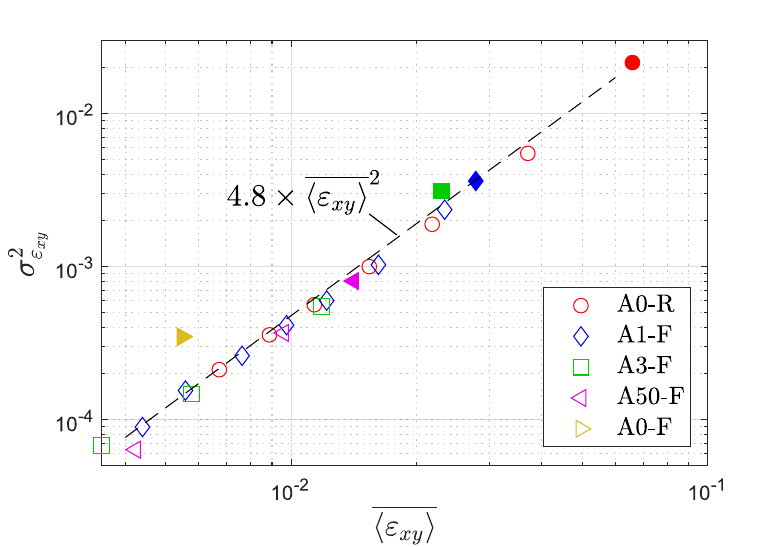}
 \caption{Strain variance as a function of mean strain for all protocols. Full colored points correspond to the last precursor before the avalanche. The filled colored symbols indicate the last precursor. \textcolor{black}{The dotted line represents a power-law fit to the experimental data, without taking into account the data associated with the A0-F protocol.}}
 \label{fig:variance}
\end{figure}
\textcolor{black}{We now examine whether the strain heterogeneities preceding the avalanche depend on the anisotropic granular structure revealed by the polar diagrams in Fig.~2(b), computed within the $10d_g\times40d_g$ region shown in Fig.~1.} The heterogeneity of the deformation field associated to precursor events can be quantified by examining the relationship between the variance of the strain $\sigma^2=\overline{\langle\cdot^2\rangle}-\overline{\langle\cdot\rangle}^2$ and its mean (see~\ref{notations} for the notations used for averaging). 
In the present study, particular care was taken to subtract the noise contribution from the strain variance, thereby allowing a broad range of strain amplitudes to be analyzed. \textcolor{black}{To estimate this noise contribution, we have measured the strain variance on an area located far away from the free surface ($|y|\in [20d_g,28d_g]$, width= $48d_g$), where the grains do not move.} Figure~\ref{fig:variance} shows that, for a given mean shear strain, the measured variance $\sigma_{\epsilon_{xy}}^2$ is independent of the preparation protocol. The data collapse around a single trend curve, with the variance increasing quadratically with the mean strain according to
$\sigma_{\epsilon_{xy}}^2=A\overline{\langle \epsilon_{xy} \rangle}^n$, where $A \approx 4.8$ and $n = 2$. This relationship may be compared to the one obtained from the heterogeneous shear provided by the exponential profile in Fig.~\ref{fig:profile}a.

The mean displacement profile is taken from paragraph~\ref{section:av_grain_motion} such that $\langle u_x(y,\theta)\rangle/d_g\approx \left(\langle u_x(0,\theta)\rangle/d_g\right)\exp(y/\lambda)$. The corresponding ensemble-averaged shear-strain profile is therefore $\langle\epsilon_{xy}\rangle(y,\theta)=\frac{1}{2}\partial\langle u_x(y,\theta)\rangle/\partial y=\epsilon_0\exp(y/\lambda)$, with $\epsilon_0=\langle u_x(0,\theta)\rangle/\left(2\lambda\right)$. With $y=0$ at the free surface and $y=-H$ at the bottom of the granular layer, the depth-averaged value over $H$ is $\overline{\langle\epsilon_{xy}\rangle}=H^{-1}\int_{-H}^{0}\langle\epsilon\rangle(y)\,\mathrm{d}y=\epsilon_0(\lambda/H)\left[1-\exp(-H/\lambda)\right]$, whereas the corresponding spatial second moment is $\overline{\langle\epsilon_{xy}\rangle^2}=H^{-1}\int_{-H}^{0}\langle\epsilon_{xy}\rangle^2(y)\,\mathrm{d}y=\epsilon_0^2(\lambda/2H)\left[1-\exp(-2H/\lambda)\right]$. 

With $H=10d_g$ and $\lambda=4d_g$, one obtains $\sigma_{\langle\epsilon_{xy}\rangle}^2\simeq0.5\overline{\langle\epsilon_{xy}\rangle}^2$. Therefore, the quadratic dependence observed experimentally in Fig.~\ref{fig:variance} is similar to that expected from an exponential mean strain profile.

 However, the measured experimental prefactor $A\approx 4.8$ is much larger than the value predicted solely from the exponential mean shear strain profile.

 Without loss of generality, the measured local shear deformation can be written $\epsilon_{xy}=\langle\epsilon_{xy}\rangle+\epsilon_{\mathrm{fluc}}$ where $\epsilon_{\mathrm{fluc}}$ denotes the local fluctuations around this profile, with $\langle\epsilon_{\mathrm{fluc}}\rangle(y)=0$, and therefore $\overline{\langle\epsilon_{\mathrm{fluc}}\rangle}=0$. The measured 
variance can then be decomposed as
$\sigma_{\epsilon_{xy}}^2=\sigma_{\langle\epsilon_{xy}\rangle}^2+\sigma_{\epsilon_{\mathrm{fluc}}}^2$ with the covariance  $\operatorname{Cov}(\langle\epsilon_{xy}\rangle,\epsilon_{\mathrm{fluc}})=0$. The term 
$\sigma_{\epsilon_{\mathrm{fluc}}}^2$ incorporates local spatial variability and accounts for $90\%$ of the measured strain variance.
\section{Discussion}
The mechanical response of a granular packing depends both on the anisotropy of its granular texture~\cite{radjai1998} and on the compatibility of this texture with the applied load~\cite{cates1998}. In the present study, during preparation of the samples with successive avalanches, the orientations of contacts and force chains become biased by the loading history, producing a mechanically anisotropic structure. 

A subsequent load is therefore accommodated more easily when it is aligned with this pre-existing structure, whereas loading in an incompatible direction requires a reorganization of the contact and force networks. Therefore, when the sample A0-F is inclined in the same direction as during avalanche preparation, the pre-existing force chains and arches are loaded predominantly in compression. This contact network can therefore accommodate the applied stress without substantial restructuring, and few or no precursors are detected. In contrast, when the sample A0-R is inclined in the opposite direction, the initial force network is no longer favorably oriented and must progressively reorganize. The reformation of force chains and arches then involve numerous localized grain rearrangements, which may participate to the observed sequence of precursors.

Interestingly, the oscillations applied during preparation can have opposite effects on precursor activity.

  As shown in Table~\ref{tab:proppacking}, samples from protocol A0-F (no oscillation) have almost no precursor while A1-F (one oscillation cycle) exhibits numerous precursors. The oscillation below the angle of repose reverses the stress applied during avalanche preparation. Our interpretation is that it is enough to alter the  force-chain network inherited from avalanche preparation and reduce its compatibility with the subsequent loading direction and promote localized rearrangements. This interpretation is supported by the polar diagrams in Fig.~\ref{fig:protocol}b: the lobe centered around $120^\circ$, observed for A0-F and associated with avalanche preparation, disappears for A1-F, while a new lobe emerges around $60^\circ$, indicating a reorientation of the granular texture induced by the single oscillation. For completeness, we report in the supplemental information Fig.S2 the evolution of the structure at each steps of the three first oscillations.

However, increasing the number of cycles can also suppress precursor activity beyond a certain number of cycles. A1-F exhibits $8.8$ precursors on average, compared with only $4.6$ for A3-F (three cycles). This decrease likely reflects progressive compaction and stabilization of the packing\cite{gravish2014,awada_2025}, which increase the number and strength of load-bearing contacts, allowing the force-chain network to accommodate the applied load with fewer rearrangements. The corresponding evolution of the polar diagrams for A3-F and A50-F in Fig.~\ref{fig:protocol}b is consistent with this progressive restructuring of the granular texture and compatible with compaction data give in the supplemental information Fig.S2. 

As shown above, the measured displacement field decays exponentially with depth according to $u_x(y,\theta)=u_x(0,\theta)\exp(y/\lambda)$, where $y$ is the depth below the free surface and $\lambda$ is the decay length. The surface displacement increases with inclination as $u_x(0,\theta) \approx u_{\mathrm{first}}\exp\left(\frac{\mu-\mu_{\mathrm{first}}}{\mu_{ra}}\right)$ with $\mu=\tan\theta$ where $\mu_{\mathrm{first}}$ marks the onset of measurable displacement and $\mu_{ra}$ controls the growth of the surface displacement. This provides directly $\text{d}y_{2.5\%}/\text{d}\mu=-\lambda/\mu_{ra}$, consistent with the slopes in Fig.~\ref{fig:profile}c. Since $\lambda$ is independent of the preparation protocol, the propagation of the deformation front in the bulk of the granular is governed by the rate of increase of the surface displacement amplitude through the parameter $\mu_{ra}$. This suggests a non-local propagation of rearrangements. 

Taken together, the results in Figs.~\ref{fig:pics}f and \ref{fig:profile}c provide insight into the front displacement involved in precursor events. This \textcolor{black}{propagation of the active zone} is characterized by $\Delta y_{2.5\%}=-\frac{\lambda}{\mu_{ra}}
\frac{\Delta\theta}{\cos^2\theta}$ where \textcolor{black}{$\Delta\theta$ is the average angular deviation to the next precursor for a precursor occurring at a given inclination angle $\theta$. }
Because the angular dependence varies only weakly, this expression can be used to estimate 
an average mobility front displacement and using the measured values of $\lambda$, $\mu_{ra}$, and $\Delta\theta$, we obtain front displacements of $0.8$, $0.9$, $2$, and $3$ grain diameters for A0-R, A1-F, A3-F and A50-F respectively. In samples such as A0-R, each new precursor expands on average the active zone by a layer approximately one grain diameter thick, whereas in A50-F, this thickness reaches approximately three grain diameters. The result suggests that precursor activity proceeds through two distinct regimes: samples with frequent precursors evolve through small, progressive increments, whereas samples with fewer precursors undergo larger, more abrupt extensions of the active zone.

Overall, our experimental results suggest that precursors develop through heterogeneous shear, characterized by a mean exponential profile and, on average, discrete increments of the active zone. In addition, Fig.~\ref{fig:variance} suggests that strain variance is dominated by strong local shear heterogeneities.

To gain insight into the large measured variance, we use an effective model where the granular medium is represented as an ensemble of shear Bagi cells. Such cells can exist in two states: a deformed ``active'' state with probability $\phi$, for which the deformation value is $\langle\epsilon_{xy}\rangle/\phi$, and an undeformed state with probability $1-\phi$. This construction preserves the exponential profile. Active cells are supposed uniformly distributed in space and independant. Within this picture, the local second moment is 
$\overline{\langle\epsilon_{xy}^2\rangle}=\overline{\langle\epsilon_{xy}\rangle^2}/\phi$, which leads to $\sigma_{\epsilon_{xy}}^2=\sigma_{\langle\epsilon_{xy}\rangle}^2/\phi+\left(1/\phi-1\right)\overline{\langle\epsilon_{xy}\rangle}^2$ or, equivalently,
$\phi=\left[1+\sigma_{\langle\epsilon_{xy}\rangle}^2\big/\overline{\langle\epsilon_{xy}\rangle}^2\right]\bigg/\left[1+\sigma_{\langle\epsilon_{xy}\rangle}^2\big/\overline{\langle\epsilon_{xy}\rangle}^2\right]$.
Using $\sigma_{\langle\epsilon_{xy}\rangle}^2\big/\overline{\langle\epsilon_{xy}\rangle}^2\simeq0.5$ and $\sigma_{\langle\epsilon_{xy}\rangle}^2\big/\overline{\langle\epsilon_{xy}\rangle}^2\simeq4.8$, we obtain $\phi\simeq0.26$. 
Within this framework, a local heterogeneity, in which the deformation is concentrated in about $1/4$ of the cells, is superimposed to the exponential mean shear-strain profile.

As a closing remark, our results raise a natural question: can this precursor description be used to forecast the avalanche (at least in the laboratory~\cite{kabla05}) or does it only describe the approach to failure, abrupt or smooth, in a ``passive'' way, without predicting it? This question is of central concern in geophysics, where considerable effort is devoted to forecasting the time of failure of catastrophic events such as rockfalls, landslides, or earthquakes~\cite{Lacroix2023,Vasseur2015}. What emerges here is that the observables we studied (surface displacement, bulk displacement, and deformation) all grow exponentially as failure is approached, unlikely to carry any predictive power for the onset of failure itself~\cite{CartwrightTaylor2020}. Moreover, our results show that the compatibility between the internal structure and the applied loading, together with its evolution through the preparation history, plays a central role in controlling precursor activity. In natural systems, however, this internal state cannot be inferred reliably from surface observations alone. 
\section*{Conclusions}
In this study, grain-scale measurements were used to characterize the heterogeneous deformation preceding avalanches in inclined granular packings. We show that precursor activity depends strongly on the preparation history and on the compatibility between the anisotropic granular structure and the subsequent loading direction. Loading along a structure inherited from previous avalanches produces few rearrangements, whereas loading in an incompatible direction promotes the progressive reorganization of the contact network and generates numerous precursors. Oscillatory preparation modifies this response by first reorienting the granular (enhancing precursors) and then, after repeated cycles, compacting and stabilizing the packing (suppressing precursors).

Despite these differences in precursor activity, the mean precursors-induced displacement exhibits a generic exponential decay with depth, with a characteristic length of approximately $4d_g$. The protocol of preparation primarily controls the growth of the displacement amplitude at the free surface and, consequently, the propagation rate of the mobility front. Samples exhibiting many precursors undergo rearrangements associated with \textcolor{black}{a front propagation} 
of approximately one grain diameter for each precursors, whereas samples exhibiting fewer precursors display more spatially coherent events associated with \textcolor{black}{a front propagation} 
extending over several grain diameters.

The shear-strain variance exhibits a common quadratic dependence on the mean strain across all preparation protocols. Its magnitude cannot be accounted for solely by the exponential mean profile. Within a simple model comprising sheared and unsheared cells, the deformation is found to be localized in approximately one fourth of the cells, providing a simple picture of the pronounced spatial heterogeneity of the strain field.

\textcolor{black}{Note that a distinctive aspect of the present work is that the precursors to granular avalanches are studied in liquid media whereas it is usually studied in air. For the large grains and low-viscosity fluid used here, fluid drag and pore-pressure effects are expected to remain negligible. Studying both the static behavior and the dynamics of precursors in immersed granular systems composed of smaller (possibly polydisperse) grains and/or more viscous fluids, where fluid--grain coupling\cite{Iverson2005} may become significant, therefore constitutes an interesting direction for future research.}

The analyses of the structure and local deformations of the granular packings reported in this paper are subject to several caveats that should be addressed before definitive conclusions can be drawn. \textcolor{black}{Wall friction may cause wall-adjacent grains to behave differently from bulk grains~\cite{TreflikBody2026} and the structure of the packing is affected by layering at the wall~\cite{gdr}.} Therefore, a full three-dimensional analysis of bulk grains is required to better understand the link between precursor activity, structure, and spatial fluctuations. This could be achieved through extensive discrete element method simulations or X-ray tomography. Moreover, in line with previous studies on the criticality of avalanches~\cite{liu1991}, finite-size effects should be considered when interpreting the increase in precursor intensity as the avalanche approaches. \textcolor{black}{Varying the cell thickness would also help determine whether the localization depth and displacement decay length depend on wall friction and confinement, as reported for granular avalanches~\cite{Jop2005}} This could help not only to relate the evolution of fluctuations to the order of the solid–fluid transition, but also to improve the prediction of avalanches and landslides in natural environments.

\section*{Author contributions}
All authors contributed equally to this work.
\section*{Conflicts of interest}
There are no conflicts to declare.
\section*{Data availability}
Data are available from the authors upon reasonable request.
\section*{Acknowledgements}
The authors gratefully acknowledge the late P. Moucheront for his valuable contributions to this work and for fruitful discussions. This work was funded by Labex MMCD.

\bibliographystyle{apsrev4-2}
\bibliography{rsc}

\begin{thebibliography}{39}%
\makeatletter
\providecommand \@ifxundefined [1]{%
 \@ifx{#1\undefined}
}%
\providecommand \@ifnum [1]{%
 \ifnum #1\expandafter \@firstoftwo
 \else \expandafter \@secondoftwo
 \fi
}%
\providecommand \@ifx [1]{%
 \ifx #1\expandafter \@firstoftwo
 \else \expandafter \@secondoftwo
 \fi
}%
\providecommand \natexlab [1]{#1}%
\providecommand \enquote  [1]{``#1''}%
\providecommand \bibnamefont  [1]{#1}%
\providecommand \bibfnamefont [1]{#1}%
\providecommand \citenamefont [1]{#1}%
\providecommand \href@noop [0]{\@secondoftwo}%
\providecommand \href [0]{\begingroup \@sanitize@url \@href}%
\providecommand \@href[1]{\@@startlink{#1}\@@href}%
\providecommand \@@href[1]{\endgroup#1\@@endlink}%
\providecommand \@sanitize@url [0]{\catcode `\\12\catcode `\$12\catcode
  `\&12\catcode `\#12\catcode `\^12\catcode `\_12\catcode `\%12\relax}%
\providecommand \@@startlink[1]{}%
\providecommand \@@endlink[0]{}%
\providecommand \url  [0]{\begingroup\@sanitize@url \@url }%
\providecommand \@url [1]{\endgroup\@href {#1}{\urlprefix }}%
\providecommand \urlprefix  [0]{URL }%
\providecommand \Eprint [0]{\href }%
\providecommand \doibase [0]{https://doi.org/}%
\providecommand \selectlanguage [0]{\@gobble}%
\providecommand \bibinfo  [0]{\@secondoftwo}%
\providecommand \bibfield  [0]{\@secondoftwo}%
\providecommand \translation [1]{[#1]}%
\providecommand \BibitemOpen [0]{}%
\providecommand \bibitemStop [0]{}%
\providecommand \bibitemNoStop [0]{.\EOS\space}%
\providecommand \EOS [0]{\spacefactor3000\relax}%
\providecommand \BibitemShut  [1]{\csname bibitem#1\endcsname}%
\let\auto@bib@innerbib\@empty
\bibitem [{\citenamefont {Kuhn}(1999)}]{Kuhn1999}%
  \BibitemOpen
  \bibfield  {author} {\bibinfo {author} {\bibfnamefont {M.~R.}\ \bibnamefont
  {Kuhn}},\ }\href@noop {} {\bibfield  {journal} {\bibinfo  {journal}
  {Mechanics of Materials}\ }\textbf {\bibinfo {volume} {31}},\ \bibinfo
  {pages} {407} (\bibinfo {year} {1999})}\BibitemShut {NoStop}%
\bibitem [{\citenamefont {Nicolas}\ \emph {et~al.}(2018)\citenamefont
  {Nicolas}, \citenamefont {Ferrero}, \citenamefont {Martens},\ and\
  \citenamefont {Barrat}}]{nicolas2018deformation}%
  \BibitemOpen
  \bibfield  {author} {\bibinfo {author} {\bibfnamefont {A.}~\bibnamefont
  {Nicolas}}, \bibinfo {author} {\bibfnamefont {E.~E.}\ \bibnamefont
  {Ferrero}}, \bibinfo {author} {\bibfnamefont {K.}~\bibnamefont {Martens}},\
  and\ \bibinfo {author} {\bibfnamefont {J.-L.}\ \bibnamefont {Barrat}},\
  }\href {https://doi.org/10.1103/RevModPhys.90.045006} {\bibfield  {journal}
  {\bibinfo  {journal} {Reviews of Modern Physics}\ }\textbf {\bibinfo {volume}
  {90}},\ \bibinfo {pages} {045006} (\bibinfo {year} {2018})}\BibitemShut
  {NoStop}%
\bibitem [{\citenamefont {Weiss}\ and\ \citenamefont
  {Marsan}(2003)}]{weiss2003threedimensional}%
  \BibitemOpen
  \bibfield  {author} {\bibinfo {author} {\bibfnamefont {J.}~\bibnamefont
  {Weiss}}\ and\ \bibinfo {author} {\bibfnamefont {D.}~\bibnamefont {Marsan}},\
  }\href {https://doi.org/10.1126/science.1079312} {\bibfield  {journal}
  {\bibinfo  {journal} {Science}\ }\textbf {\bibinfo {volume} {299}},\ \bibinfo
  {pages} {89} (\bibinfo {year} {2003})}\BibitemShut {NoStop}%
\bibitem [{\citenamefont {Gopal}\ and\ \citenamefont
  {Durian}(1995)}]{gopal1995nonlinear}%
  \BibitemOpen
  \bibfield  {author} {\bibinfo {author} {\bibfnamefont {A.~D.}\ \bibnamefont
  {Gopal}}\ and\ \bibinfo {author} {\bibfnamefont {D.~J.}\ \bibnamefont
  {Durian}},\ }\href {https://doi.org/10.1103/PhysRevLett.75.2610} {\bibfield
  {journal} {\bibinfo  {journal} {Physical Review Letters}\ }\textbf {\bibinfo
  {volume} {75}},\ \bibinfo {pages} {2610} (\bibinfo {year}
  {1995})}\BibitemShut {NoStop}%
\bibitem [{\citenamefont {Weeks}\ \emph {et~al.}(2000)\citenamefont {Weeks},
  \citenamefont {Crocker}, \citenamefont {Levitt}, \citenamefont {Schofield},\
  and\ \citenamefont {Weitz}}]{weeks}%
  \BibitemOpen
  \bibfield  {author} {\bibinfo {author} {\bibfnamefont {E.~R.}\ \bibnamefont
  {Weeks}}, \bibinfo {author} {\bibfnamefont {J.~C.}\ \bibnamefont {Crocker}},
  \bibinfo {author} {\bibfnamefont {A.~C.}\ \bibnamefont {Levitt}}, \bibinfo
  {author} {\bibfnamefont {A.}~\bibnamefont {Schofield}},\ and\ \bibinfo
  {author} {\bibfnamefont {D.~A.}\ \bibnamefont {Weitz}},\ }\href@noop {}
  {\bibfield  {journal} {\bibinfo  {journal} {Science}\ }\textbf {\bibinfo
  {volume} {287}},\ \bibinfo {pages} {627} (\bibinfo {year}
  {2000})}\BibitemShut {NoStop}%
\bibitem [{\citenamefont {Buisson}\ \emph {et~al.}(2003)\citenamefont
  {Buisson}, \citenamefont {Bellon},\ and\ \citenamefont
  {Ciliberto}}]{buisson2003intermittency}%
  \BibitemOpen
  \bibfield  {author} {\bibinfo {author} {\bibfnamefont {L.}~\bibnamefont
  {Buisson}}, \bibinfo {author} {\bibfnamefont {L.}~\bibnamefont {Bellon}},\
  and\ \bibinfo {author} {\bibfnamefont {S.}~\bibnamefont {Ciliberto}},\ }\href
  {https://doi.org/10.1088/0953-8984/15/11/332} {\bibfield  {journal} {\bibinfo
   {journal} {Journal of Physics: Condensed Matter}\ }\textbf {\bibinfo
  {volume} {15}},\ \bibinfo {pages} {S1163} (\bibinfo {year}
  {2003})}\BibitemShut {NoStop}%
\bibitem [{\citenamefont {Baumberger}\ and\ \citenamefont
  {Caroli}(2006)}]{carobaumb}%
  \BibitemOpen
  \bibfield  {author} {\bibinfo {author} {\bibfnamefont {T.}~\bibnamefont
  {Baumberger}}\ and\ \bibinfo {author} {\bibfnamefont {C.}~\bibnamefont
  {Caroli}},\ }\href@noop {} {\bibfield  {journal} {\bibinfo  {journal} {Adv.\
  Phys.}\ }\textbf {\bibinfo {volume} {55}},\ \bibinfo {pages} {279} (\bibinfo
  {year} {2006})}\BibitemShut {NoStop}%
\bibitem [{\citenamefont {Staron}\ \emph {et~al.}(2002)\citenamefont {Staron},
  \citenamefont {Vilotte},\ and\ \citenamefont {Radjai}}]{vilotte}%
  \BibitemOpen
  \bibfield  {author} {\bibinfo {author} {\bibfnamefont {L.}~\bibnamefont
  {Staron}}, \bibinfo {author} {\bibfnamefont {J.}~\bibnamefont {Vilotte}},\
  and\ \bibinfo {author} {\bibfnamefont {F.}~\bibnamefont {Radjai}},\
  }\href@noop {} {\bibfield  {journal} {\bibinfo  {journal} {Phys. Rev. Lett.}\
  }\textbf {\bibinfo {volume} {89}},\ \bibinfo {pages} {204302} (\bibinfo
  {year} {2002})}\BibitemShut {NoStop}%
\bibitem [{\citenamefont {Scholz}(2019)}]{scholz2019}%
  \BibitemOpen
  \bibfield  {author} {\bibinfo {author} {\bibfnamefont {C.~H.}\ \bibnamefont
  {Scholz}},\ }\href {https://doi.org/10.1017/9781316681473} {\emph {\bibinfo
  {title} {The Mechanics of Earthquakes and Faulting}}},\ \bibinfo {edition}
  {3rd}\ ed.\ (\bibinfo  {publisher} {Cambridge University Press},\ \bibinfo
  {address} {Cambridge},\ \bibinfo {year} {2019})\BibitemShut {NoStop}%
\bibitem [{\citenamefont {Petley}\ \emph {et~al.}(2002)\citenamefont {Petley},
  \citenamefont {Bulmer},\ and\ \citenamefont {Murphy}}]{petley2002}%
  \BibitemOpen
  \bibfield  {author} {\bibinfo {author} {\bibfnamefont {D.~N.}\ \bibnamefont
  {Petley}}, \bibinfo {author} {\bibfnamefont {M.~H.}\ \bibnamefont {Bulmer}},\
  and\ \bibinfo {author} {\bibfnamefont {W.}~\bibnamefont {Murphy}},\
  }\href@noop {} {\bibfield  {journal} {\bibinfo  {journal} {Geology}\ }\textbf
  {\bibinfo {volume} {30}},\ \bibinfo {pages} {719} (\bibinfo {year}
  {2002})}\BibitemShut {NoStop}%
\bibitem [{\citenamefont {Amitrano}\ \emph {et~al.}(2005)\citenamefont
  {Amitrano}, \citenamefont {Grasso},\ and\ \citenamefont
  {Senfaute}}]{amitrano2005}%
  \BibitemOpen
  \bibfield  {author} {\bibinfo {author} {\bibfnamefont {D.}~\bibnamefont
  {Amitrano}}, \bibinfo {author} {\bibfnamefont {J.-R.}\ \bibnamefont
  {Grasso}},\ and\ \bibinfo {author} {\bibfnamefont {G.}~\bibnamefont
  {Senfaute}},\ }\href@noop {} {\bibfield  {journal} {\bibinfo  {journal}
  {Geophys. Res. Lett.}\ }\textbf {\bibinfo {volume} {32}},\ \bibinfo {pages}
  {L08314} (\bibinfo {year} {2005})}\BibitemShut {NoStop}%
\bibitem [{\citenamefont {Nasuno}\ \emph {et~al.}(1998)\citenamefont {Nasuno},
  \citenamefont {Kudrolli}, \citenamefont {Bak},\ and\ \citenamefont
  {Gollub}}]{nasuno1998}%
  \BibitemOpen
  \bibfield  {author} {\bibinfo {author} {\bibfnamefont {S.}~\bibnamefont
  {Nasuno}}, \bibinfo {author} {\bibfnamefont {A.}~\bibnamefont {Kudrolli}},
  \bibinfo {author} {\bibfnamefont {A.}~\bibnamefont {Bak}},\ and\ \bibinfo
  {author} {\bibfnamefont {J.~P.}\ \bibnamefont {Gollub}},\ }\href
  {https://doi.org/10.1103/PhysRevE.58.2161} {\bibfield  {journal} {\bibinfo
  {journal} {Phys. Rev. E}\ }\textbf {\bibinfo {volume} {58}},\ \bibinfo
  {pages} {2161} (\bibinfo {year} {1998})}\BibitemShut {NoStop}%
\bibitem [{\citenamefont {Ferdowsi}\ \emph {et~al.}(2013)\citenamefont
  {Ferdowsi}, \citenamefont {Griffa}, \citenamefont {Guyer}, \citenamefont
  {Johnson}, \citenamefont {Marone},\ and\ \citenamefont
  {Carmeliet}}]{ferdowsi2013}%
  \BibitemOpen
  \bibfield  {author} {\bibinfo {author} {\bibfnamefont {B.}~\bibnamefont
  {Ferdowsi}}, \bibinfo {author} {\bibfnamefont {M.}~\bibnamefont {Griffa}},
  \bibinfo {author} {\bibfnamefont {R.~A.}\ \bibnamefont {Guyer}}, \bibinfo
  {author} {\bibfnamefont {P.~A.}\ \bibnamefont {Johnson}}, \bibinfo {author}
  {\bibfnamefont {C.}~\bibnamefont {Marone}},\ and\ \bibinfo {author}
  {\bibfnamefont {J.}~\bibnamefont {Carmeliet}},\ }\href
  {https://doi.org/10.1002/grl.50813} {\bibfield  {journal} {\bibinfo
  {journal} {Geophys. Res. Lett.}\ }\textbf {\bibinfo {volume} {40}},\ \bibinfo
  {pages} {4194} (\bibinfo {year} {2013})}\BibitemShut {NoStop}%
\bibitem [{\citenamefont {Nerone}\ \emph
  {et~al.}(2003{\natexlab{a}})\citenamefont {Nerone}, \citenamefont {Aguirre},
  \citenamefont {Calvo}, \citenamefont {Bideau},\ and\ \citenamefont
  {Ippolito}}]{nerone2003}%
  \BibitemOpen
  \bibfield  {author} {\bibinfo {author} {\bibfnamefont {N.}~\bibnamefont
  {Nerone}}, \bibinfo {author} {\bibfnamefont {M.~A.}\ \bibnamefont {Aguirre}},
  \bibinfo {author} {\bibfnamefont {A.}~\bibnamefont {Calvo}}, \bibinfo
  {author} {\bibfnamefont {D.}~\bibnamefont {Bideau}},\ and\ \bibinfo {author}
  {\bibfnamefont {I.}~\bibnamefont {Ippolito}},\ }\href@noop {} {\bibfield
  {journal} {\bibinfo  {journal} {Phys. Rev. E}\ }\textbf {\bibinfo {volume}
  {67}},\ \bibinfo {pages} {011302} (\bibinfo {year}
  {2003}{\natexlab{a}})}\BibitemShut {NoStop}%
\bibitem [{\citenamefont {Amon}\ \emph {et~al.}(2013)\citenamefont {Amon},
  \citenamefont {Bertoni}, \citenamefont {Crassous},\ and\ \citenamefont
  {Clément}}]{amon2013}%
  \BibitemOpen
  \bibfield  {author} {\bibinfo {author} {\bibfnamefont {A.}~\bibnamefont
  {Amon}}, \bibinfo {author} {\bibfnamefont {R.}~\bibnamefont {Bertoni}},
  \bibinfo {author} {\bibfnamefont {J.}~\bibnamefont {Crassous}},\ and\
  \bibinfo {author} {\bibfnamefont {E.}~\bibnamefont {Clément}},\ }\href@noop
  {} {\bibfield  {journal} {\bibinfo  {journal} {Phys. Rev. E}\ }\textbf
  {\bibinfo {volume} {87}},\ \bibinfo {pages} {012204} (\bibinfo {year}
  {2013})}\BibitemShut {NoStop}%
\bibitem [{\citenamefont {Gravish}\ and\ \citenamefont
  {Goldman}(2014)}]{gravish2014}%
  \BibitemOpen
  \bibfield  {author} {\bibinfo {author} {\bibfnamefont {N.}~\bibnamefont
  {Gravish}}\ and\ \bibinfo {author} {\bibfnamefont {D.~I.}\ \bibnamefont
  {Goldman}},\ }\href@noop {} {\bibfield  {journal} {\bibinfo  {journal} {Phys.
  Rev. E}\ }\textbf {\bibinfo {volume} {90}},\ \bibinfo {pages} {032202}
  (\bibinfo {year} {2014})}\BibitemShut {NoStop}%
\bibitem [{\citenamefont {Oger}\ \emph {et~al.}(2007)\citenamefont {Oger},
  \citenamefont {Ippolito},\ and\ \citenamefont {Vidales}}]{oger2007disorder}%
  \BibitemOpen
  \bibfield  {author} {\bibinfo {author} {\bibfnamefont {L.}~\bibnamefont
  {Oger}}, \bibinfo {author} {\bibfnamefont {I.}~\bibnamefont {Ippolito}},\
  and\ \bibinfo {author} {\bibfnamefont {A.~M.}\ \bibnamefont {Vidales}},\
  }\href {https://doi.org/10.1007/s10035-007-0040-8} {\bibfield  {journal}
  {\bibinfo  {journal} {Granular Matter}\ }\textbf {\bibinfo {volume} {9}},\
  \bibinfo {pages} {267} (\bibinfo {year} {2007})}\BibitemShut {NoStop}%
\bibitem [{\citenamefont {Kiesgen~de Richter}\ \emph
  {et~al.}(2012)\citenamefont {Kiesgen~de Richter}, \citenamefont {Le~Caër},\
  and\ \citenamefont {Delannay}}]{kiesgen12}%
  \BibitemOpen
  \bibfield  {author} {\bibinfo {author} {\bibfnamefont {S.}~\bibnamefont
  {Kiesgen~de Richter}}, \bibinfo {author} {\bibfnamefont {G.}~\bibnamefont
  {Le~Caër}},\ and\ \bibinfo {author} {\bibfnamefont {R.}~\bibnamefont
  {Delannay}},\ }\href@noop {} {\bibfield  {journal} {\bibinfo  {journal}
  {Journal of Statistical Mechanics: Theory and Experiment}\ }\textbf {\bibinfo
  {volume} {2012}},\ \bibinfo {pages} {P04013} (\bibinfo {year}
  {2012})}\BibitemShut {NoStop}%
\bibitem [{\citenamefont {Evesque}\ \emph {et~al.}(1993)\citenamefont
  {Evesque}, \citenamefont {Fargeix}, \citenamefont {Habib}, \citenamefont
  {Luong},\ and\ \citenamefont {Porion}}]{evesque1993piledensity}%
  \BibitemOpen
  \bibfield  {author} {\bibinfo {author} {\bibfnamefont {P.}~\bibnamefont
  {Evesque}}, \bibinfo {author} {\bibfnamefont {D.}~\bibnamefont {Fargeix}},
  \bibinfo {author} {\bibfnamefont {P.}~\bibnamefont {Habib}}, \bibinfo
  {author} {\bibfnamefont {M.~P.}\ \bibnamefont {Luong}},\ and\ \bibinfo
  {author} {\bibfnamefont {P.}~\bibnamefont {Porion}},\ }\href
  {https://doi.org/10.1103/PhysRevE.47.2326} {\bibfield  {journal} {\bibinfo
  {journal} {Physical Review E}\ }\textbf {\bibinfo {volume} {47}},\ \bibinfo
  {pages} {2326} (\bibinfo {year} {1993})}\BibitemShut {NoStop}%
\bibitem [{\citenamefont {Schmidt}\ \emph {et~al.}(2018)\citenamefont
  {Schmidt}, \citenamefont {Weigert}, \citenamefont {Broaddus},\ and\
  \citenamefont {Myers}}]{stardist}%
  \BibitemOpen
  \bibfield  {author} {\bibinfo {author} {\bibfnamefont {U.}~\bibnamefont
  {Schmidt}}, \bibinfo {author} {\bibfnamefont {M.}~\bibnamefont {Weigert}},
  \bibinfo {author} {\bibfnamefont {C.}~\bibnamefont {Broaddus}},\ and\
  \bibinfo {author} {\bibfnamefont {G.}~\bibnamefont {Myers}},\ }\href@noop {}
  {\bibfield  {journal} {\bibinfo  {journal} {CoRR}\ }\textbf {\bibinfo
  {volume} {abs/1806.03535}},\ \bibinfo {pages} {arXiv:1806.03535} (\bibinfo
  {year} {2018})}\BibitemShut {NoStop}%
\bibitem [{\citenamefont {Awada}\ \emph
  {et~al.}(2025{\natexlab{a}})\citenamefont {Awada}, \citenamefont {Bornert},
  \citenamefont {Langlois},\ and\ \citenamefont {Léopoldès}}]{awada_2025}%
  \BibitemOpen
  \bibfield  {author} {\bibinfo {author} {\bibfnamefont {I.}~\bibnamefont
  {Awada}}, \bibinfo {author} {\bibfnamefont {M.}~\bibnamefont {Bornert}},
  \bibinfo {author} {\bibfnamefont {V.}~\bibnamefont {Langlois}},\ and\
  \bibinfo {author} {\bibfnamefont {J.}~\bibnamefont {Léopoldès}},\ }\href
  {https://www.arxiv.org/abs/2503.17010} {\bibfield  {journal} {\bibinfo
  {journal} {preprint arXiv:2503.17010}\ } (\bibinfo {year}
  {2025}{\natexlab{a}})}\BibitemShut {NoStop}%
\bibitem [{\citenamefont {Bagi}(1996)}]{Bagi1996}%
  \BibitemOpen
  \bibfield  {author} {\bibinfo {author} {\bibfnamefont {K.}~\bibnamefont
  {Bagi}},\ }\href@noop {} {\bibfield  {journal} {\bibinfo  {journal}
  {Mechanics of Materials}\ }\textbf {\bibinfo {volume} {22}},\ \bibinfo
  {pages} {165} (\bibinfo {year} {1996})}\BibitemShut {NoStop}%
\bibitem [{\citenamefont {Awada}\ \emph
  {et~al.}(2025{\natexlab{b}})\citenamefont {Awada}, \citenamefont {Bornert},
  \citenamefont {Langlois},\ and\ \citenamefont
  {L{\'e}opold{\`e}s}}]{Awada2025}%
  \BibitemOpen
  \bibfield  {author} {\bibinfo {author} {\bibfnamefont {I.}~\bibnamefont
  {Awada}}, \bibinfo {author} {\bibfnamefont {M.}~\bibnamefont {Bornert}},
  \bibinfo {author} {\bibfnamefont {V.}~\bibnamefont {Langlois}},\ and\
  \bibinfo {author} {\bibfnamefont {J.}~\bibnamefont {L{\'e}opold{\`e}s}},\
  }\href@noop {} {\bibfield  {journal} {\bibinfo  {journal} {Eur. Phys. J. E}\
  }\textbf {\bibinfo {volume} {48}},\ \bibinfo {pages} {53} (\bibinfo {year}
  {2025}{\natexlab{b}})}\BibitemShut {NoStop}%
\bibitem [{\citenamefont {Tsai}\ and\ \citenamefont
  {Gollub}(2004)}]{TsaiGollub2004}%
  \BibitemOpen
  \bibfield  {author} {\bibinfo {author} {\bibfnamefont {J.-C.}\ \bibnamefont
  {Tsai}}\ and\ \bibinfo {author} {\bibfnamefont {J.~P.}\ \bibnamefont
  {Gollub}},\ }\href {https://doi.org/10.1103/PhysRevE.70.031303} {\bibfield
  {journal} {\bibinfo  {journal} {Physical Review E}\ }\textbf {\bibinfo
  {volume} {70}},\ \bibinfo {pages} {031303} (\bibinfo {year}
  {2004})}\BibitemShut {NoStop}%
\bibitem [{\citenamefont {Wegner}\ \emph {et~al.}(2014)\citenamefont {Wegner},
  \citenamefont {Stannarius}, \citenamefont {Boese}, \citenamefont {Rose},
  \citenamefont {Szab{\'o}}, \citenamefont {Somfai},\ and\ \citenamefont
  {B{\"o}rzs{\"o}nyi}}]{WegnerEtAl2014}%
  \BibitemOpen
  \bibfield  {author} {\bibinfo {author} {\bibfnamefont {S.}~\bibnamefont
  {Wegner}}, \bibinfo {author} {\bibfnamefont {R.}~\bibnamefont {Stannarius}},
  \bibinfo {author} {\bibfnamefont {A.}~\bibnamefont {Boese}}, \bibinfo
  {author} {\bibfnamefont {G.}~\bibnamefont {Rose}}, \bibinfo {author}
  {\bibfnamefont {B.}~\bibnamefont {Szab{\'o}}}, \bibinfo {author}
  {\bibfnamefont {E.}~\bibnamefont {Somfai}},\ and\ \bibinfo {author}
  {\bibfnamefont {T.}~\bibnamefont {B{\"o}rzs{\"o}nyi}},\ }\href
  {https://doi.org/10.1039/C4SM00838C} {\bibfield  {journal} {\bibinfo
  {journal} {Soft Matter}\ }\textbf {\bibinfo {volume} {10}},\ \bibinfo {pages}
  {5157} (\bibinfo {year} {2014})}\BibitemShut {NoStop}%
\bibitem [{\citenamefont {Nerone}\ \emph {et~al.}(2000)\citenamefont {Nerone},
  \citenamefont {Aguirre}, \citenamefont {Calvo}, \citenamefont {Ippolito},\
  and\ \citenamefont {Bideau}}]{nerone2000}%
  \BibitemOpen
  \bibfield  {author} {\bibinfo {author} {\bibfnamefont {N.}~\bibnamefont
  {Nerone}}, \bibinfo {author} {\bibfnamefont {M.~A.}\ \bibnamefont {Aguirre}},
  \bibinfo {author} {\bibfnamefont {A.}~\bibnamefont {Calvo}}, \bibinfo
  {author} {\bibfnamefont {I.}~\bibnamefont {Ippolito}},\ and\ \bibinfo
  {author} {\bibfnamefont {D.}~\bibnamefont {Bideau}},\ }\href@noop {}
  {\bibfield  {journal} {\bibinfo  {journal} {Physica A: Statistical Mechanics
  and its Applications}\ }\textbf {\bibinfo {volume} {283}},\ \bibinfo {pages}
  {218} (\bibinfo {year} {2000})}\BibitemShut {NoStop}%
\bibitem [{\citenamefont {Nerone}\ \emph
  {et~al.}(2003{\natexlab{b}})\citenamefont {Nerone}, \citenamefont {Aguirre},
  \citenamefont {Calvo}, \citenamefont {Bideau},\ and\ \citenamefont
  {Ippolito}}]{nerone}%
  \BibitemOpen
  \bibfield  {author} {\bibinfo {author} {\bibfnamefont {N.}~\bibnamefont
  {Nerone}}, \bibinfo {author} {\bibfnamefont {M.~A.}\ \bibnamefont {Aguirre}},
  \bibinfo {author} {\bibfnamefont {A.}~\bibnamefont {Calvo}}, \bibinfo
  {author} {\bibfnamefont {D.}~\bibnamefont {Bideau}},\ and\ \bibinfo {author}
  {\bibfnamefont {I.}~\bibnamefont {Ippolito}},\ }\href@noop {} {\bibfield
  {journal} {\bibinfo  {journal} {Phys. Rev. E}\ }\textbf {\bibinfo {volume}
  {67}},\ \bibinfo {pages} {011302} (\bibinfo {year}
  {2003}{\natexlab{b}})}\BibitemShut {NoStop}%
\bibitem [{\citenamefont {Munro}(1997)}]{Munro1997}%
  \BibitemOpen
  \bibfield  {author} {\bibinfo {author} {\bibfnamefont {R.~G.}\ \bibnamefont
  {Munro}},\ }\href {https://doi.org/10.1111/j.1151-2916.1997.tb03074.x}
  {\bibfield  {journal} {\bibinfo  {journal} {Journal of the American Ceramic
  Society}\ }\textbf {\bibinfo {volume} {80}},\ \bibinfo {pages} {1919}
  (\bibinfo {year} {1997})}\BibitemShut {NoStop}%
\bibitem [{\citenamefont {Kabla}\ \emph {et~al.}(2005)\citenamefont {Kabla},
  \citenamefont {Debrégeas}, \citenamefont {di~Meglio},\ and\ \citenamefont
  {Senden}}]{kabla05}%
  \BibitemOpen
  \bibfield  {author} {\bibinfo {author} {\bibfnamefont {A.}~\bibnamefont
  {Kabla}}, \bibinfo {author} {\bibfnamefont {G.}~\bibnamefont {Debrégeas}},
  \bibinfo {author} {\bibfnamefont {J.-M.}\ \bibnamefont {di~Meglio}},\ and\
  \bibinfo {author} {\bibfnamefont {T.}~\bibnamefont {Senden}},\ }\href@noop {}
  {\bibfield  {journal} {\bibinfo  {journal} {EPL}\ }\textbf {\bibinfo {volume}
  {71}},\ \bibinfo {pages} {932} (\bibinfo {year} {2005})}\BibitemShut
  {NoStop}%
\bibitem [{\citenamefont {Radjai}\ \emph {et~al.}(1998)\citenamefont {Radjai},
  \citenamefont {Wolf}, \citenamefont {Jean},\ and\ \citenamefont
  {Moreau}}]{radjai1998}%
  \BibitemOpen
  \bibfield  {author} {\bibinfo {author} {\bibfnamefont {F.}~\bibnamefont
  {Radjai}}, \bibinfo {author} {\bibfnamefont {D.}~\bibnamefont {Wolf}},
  \bibinfo {author} {\bibfnamefont {M.}~\bibnamefont {Jean}},\ and\ \bibinfo
  {author} {\bibfnamefont {J.-J.}\ \bibnamefont {Moreau}},\ }\href@noop {}
  {\bibfield  {journal} {\bibinfo  {journal} {Phys. Rev. Letters}\ }\textbf
  {\bibinfo {volume} {80}},\ \bibinfo {pages} {61} (\bibinfo {year}
  {1998})}\BibitemShut {NoStop}%
\bibitem [{\citenamefont {Cates}\ \emph {et~al.}(1998)\citenamefont {Cates},
  \citenamefont {Wittmer}, \citenamefont {Bouchaud},\ and\ \citenamefont
  {Claudin}}]{cates1998}%
  \BibitemOpen
  \bibfield  {author} {\bibinfo {author} {\bibfnamefont {M.~E.}\ \bibnamefont
  {Cates}}, \bibinfo {author} {\bibfnamefont {J.~P.}\ \bibnamefont {Wittmer}},
  \bibinfo {author} {\bibfnamefont {J.-P.}\ \bibnamefont {Bouchaud}},\ and\
  \bibinfo {author} {\bibfnamefont {P.}~\bibnamefont {Claudin}},\ }\href@noop
  {} {\bibfield  {journal} {\bibinfo  {journal} {Phys. Rev. Lett.}\ }\textbf
  {\bibinfo {volume} {81}},\ \bibinfo {pages} {1841} (\bibinfo {year}
  {1998})}\BibitemShut {NoStop}%
\bibitem [{\citenamefont {Lacroix}\ \emph {et~al.}(2023)\citenamefont
  {Lacroix}, \citenamefont {Huanca}, \citenamefont {Albinez},\ and\
  \citenamefont {Taipe}}]{Lacroix2023}%
  \BibitemOpen
  \bibfield  {author} {\bibinfo {author} {\bibfnamefont {P.}~\bibnamefont
  {Lacroix}}, \bibinfo {author} {\bibfnamefont {J.}~\bibnamefont {Huanca}},
  \bibinfo {author} {\bibfnamefont {L.}~\bibnamefont {Albinez}},\ and\ \bibinfo
  {author} {\bibfnamefont {E.}~\bibnamefont {Taipe}},\ }\href
  {https://doi.org/10.1029/2023GL105413} {\bibfield  {journal} {\bibinfo
  {journal} {Geophysical Research Letters}\ }\textbf {\bibinfo {volume} {50}},\
  \bibinfo {pages} {e2023GL105413} (\bibinfo {year} {2023})}\BibitemShut
  {NoStop}%
\bibitem [{\citenamefont {Vasseur}\ \emph {et~al.}(2015)\citenamefont
  {Vasseur}, \citenamefont {Wadsworth}, \citenamefont {Lavall{\'e}e},
  \citenamefont {Bell}, \citenamefont {Main},\ and\ \citenamefont
  {Dingwell}}]{Vasseur2015}%
  \BibitemOpen
  \bibfield  {author} {\bibinfo {author} {\bibfnamefont {J.}~\bibnamefont
  {Vasseur}}, \bibinfo {author} {\bibfnamefont {F.~B.}\ \bibnamefont
  {Wadsworth}}, \bibinfo {author} {\bibfnamefont {Y.}~\bibnamefont
  {Lavall{\'e}e}}, \bibinfo {author} {\bibfnamefont {A.~F.}\ \bibnamefont
  {Bell}}, \bibinfo {author} {\bibfnamefont {I.~G.}\ \bibnamefont {Main}},\
  and\ \bibinfo {author} {\bibfnamefont {D.~B.}\ \bibnamefont {Dingwell}},\
  }\href {https://doi.org/10.1038/srep13259} {\bibfield  {journal} {\bibinfo
  {journal} {Scientific Reports}\ }\textbf {\bibinfo {volume} {5}},\ \bibinfo
  {pages} {13259} (\bibinfo {year} {2015})}\BibitemShut {NoStop}%
\bibitem [{\citenamefont {Cartwright-Taylor}\ \emph {et~al.}(2020)\citenamefont
  {Cartwright-Taylor}, \citenamefont {Main}, \citenamefont {Butler},
  \citenamefont {Fusseis}, \citenamefont {Flynn},\ and\ \citenamefont
  {King}}]{CartwrightTaylor2020}%
  \BibitemOpen
  \bibfield  {author} {\bibinfo {author} {\bibfnamefont {A.}~\bibnamefont
  {Cartwright-Taylor}}, \bibinfo {author} {\bibfnamefont {I.~G.}\ \bibnamefont
  {Main}}, \bibinfo {author} {\bibfnamefont {I.~B.}\ \bibnamefont {Butler}},
  \bibinfo {author} {\bibfnamefont {F.}~\bibnamefont {Fusseis}}, \bibinfo
  {author} {\bibfnamefont {M.}~\bibnamefont {Flynn}},\ and\ \bibinfo {author}
  {\bibfnamefont {A.}~\bibnamefont {King}},\ }\href
  {https://doi.org/10.1029/2020JB019642} {\bibfield  {journal} {\bibinfo
  {journal} {Journal of Geophysical Research: Solid Earth}\ }\textbf {\bibinfo
  {volume} {125}},\ \bibinfo {pages} {e2020JB019642} (\bibinfo {year}
  {2020})}\BibitemShut {NoStop}%
\bibitem [{\citenamefont {Iverson}(2005)}]{Iverson2005}%
  \BibitemOpen
  \bibfield  {author} {\bibinfo {author} {\bibfnamefont {R.~M.}\ \bibnamefont
  {Iverson}},\ }\href {https://doi.org/10.1029/2004JF000268} {\bibfield
  {journal} {\bibinfo  {journal} {Journal of Geophysical Research: Earth
  Surface}\ }\textbf {\bibinfo {volume} {110}},\ \bibinfo {pages} {F02015}
  (\bibinfo {year} {2005})}\BibitemShut {NoStop}%
\bibitem [{\citenamefont {Treflik-Body}\ \emph {et~al.}(2026)\citenamefont
  {Treflik-Body}, \citenamefont {Omer}, \citenamefont {Talesnick},
  \citenamefont {Steel}, \citenamefont {Mulligan},\ and\ \citenamefont
  {Take}}]{TreflikBody2026}%
  \BibitemOpen
  \bibfield  {author} {\bibinfo {author} {\bibfnamefont {E.}~\bibnamefont
  {Treflik-Body}}, \bibinfo {author} {\bibfnamefont {I.}~\bibnamefont {Omer}},
  \bibinfo {author} {\bibfnamefont {M.}~\bibnamefont {Talesnick}}, \bibinfo
  {author} {\bibfnamefont {E.}~\bibnamefont {Steel}}, \bibinfo {author}
  {\bibfnamefont {R.~P.}\ \bibnamefont {Mulligan}},\ and\ \bibinfo {author}
  {\bibfnamefont {W.~A.}\ \bibnamefont {Take}},\ }\bibfield  {journal}
  {\bibinfo  {journal} {Canadian Geotechnical Journal}\ }\href
  {https://doi.org/10.1139/cgj-2026-0016} {10.1139/cgj-2026-0016} (\bibinfo
  {year} {2026})\BibitemShut {NoStop}%
\bibitem [{\citenamefont {{GDR MiDi}}(2004)}]{gdr}%
  \BibitemOpen
  \bibfield  {author} {\bibinfo {author} {\bibnamefont {{GDR MiDi}}},\ }\href
  {https://doi.org/10.1140/epje/i2003-10153-0} {\bibfield  {journal} {\bibinfo
  {journal} {The European Physical Journal E}\ }\textbf {\bibinfo {volume}
  {14}},\ \bibinfo {pages} {341} (\bibinfo {year} {2004})}\BibitemShut
  {NoStop}%
\bibitem [{\citenamefont {Liu}\ \emph {et~al.}(1991)\citenamefont {Liu},
  \citenamefont {Jaeger},\ and\ \citenamefont {Nagel}}]{liu1991}%
  \BibitemOpen
  \bibfield  {author} {\bibinfo {author} {\bibfnamefont {C.-H.}\ \bibnamefont
  {Liu}}, \bibinfo {author} {\bibfnamefont {H.~M.}\ \bibnamefont {Jaeger}},\
  and\ \bibinfo {author} {\bibfnamefont {S.~R.}\ \bibnamefont {Nagel}},\ }\href
  {https://doi.org/10.1103/PhysRevA.43.7091} {\bibfield  {journal} {\bibinfo
  {journal} {Phys. Rev. A}\ }\textbf {\bibinfo {volume} {43}},\ \bibinfo
  {pages} {7091} (\bibinfo {year} {1991})}\BibitemShut {NoStop}%
\bibitem [{\citenamefont {Jop}\ \emph {et~al.}(2005)\citenamefont {Jop},
  \citenamefont {Forterre},\ and\ \citenamefont {Pouliquen}}]{Jop2005}%
  \BibitemOpen
  \bibfield  {author} {\bibinfo {author} {\bibfnamefont {P.}~\bibnamefont
  {Jop}}, \bibinfo {author} {\bibfnamefont {Y.}~\bibnamefont {Forterre}},\ and\
  \bibinfo {author} {\bibfnamefont {O.}~\bibnamefont {Pouliquen}},\ }\href@noop
  {} {\bibfield  {journal} {\bibinfo  {journal} {J. Fluid Mech.}\ }\textbf
  {\bibinfo {volume} {541}},\ \bibinfo {pages} {167} (\bibinfo {year}
  {2005})}\BibitemShut {NoStop}%
\end{thebibliography}%

\clearpage
\section*{Supplemental Material}
\begin{center}
\textbf{Granular structure and heterogeneous deformation preceding avalanches}\\[0.5em]
Ibrahim Awada, Michel Bornert, Vincent Langlois, and Julien L\'eopold\`es
\end{center}
\vspace{1em}

\setcounter{figure}{0}
\renewcommand{\thefigure}{S\arabic{figure}}
\setcounter{table}{0}
\renewcommand{\thetable}{S\arabic{table}}
\setcounter{equation}{0}
\renewcommand{\theequation}{S\arabic{equation}}

\section*{Effect of the oscillations on the apparent solid fraction.}

In this Supplemental Material, we present additional results on the effect of the cell oscillations applied during sample preparation on the apparent solid fraction and on the local grain structure characterized through polar diagrams of neighboring-grain orientations.

\begin{figure}[htbp]
    \centering
    \includegraphics[width=0.8\linewidth]{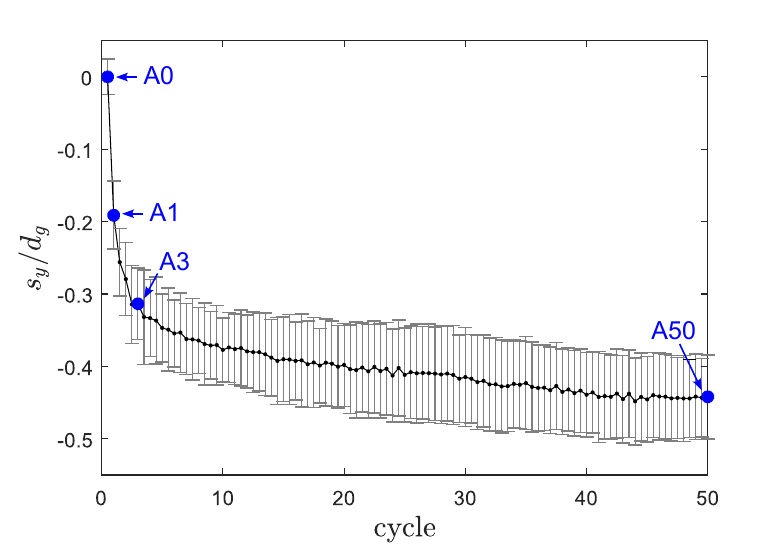}
    \caption{Displacement of the free surface $s_y$ as a function of the number of oscillations applied during the preparation protocol shown in Fig.~\ref{fig:protocol}a. The position of the free surface at position $b_1$ (Fig.\ref{fig:protocol}a) is taken as the reference, and $s_y$ is normalized by the diameter of the grains. A0, A1, A3 and A50 refer to the displacement of the free surface at the end of the preparation for each corresponding protocol. Error bars correspond to the standard deviation computed from independent repetitions of the experiment.
    }
    \label{fig:freesurf}
\end{figure}

As shown in a previous study\cite{awada_2025}, oscillations lead to an increase in the apparent solid fraction. This compaction can be monitored by observing the displacement of the free surface $s_y$, as shown in Fig.~\ref{fig:freesurf}.
The gradual lowering of the free-surface level reflects the compaction of the granular packing. The corresponding relative variation in the global packing fraction can be estimated as $\Delta\phi/\phi_0 \simeq s_y/y_a$, where $\phi_0 \simeq 0.61$ denotes the initial packing fraction and $y_a$ is the thickness of the sheared zone during avalanche flow. After 50 cycles, the packing fraction of A50-F sample has increased to $\phi \simeq 0.63$, corresponding to a more symmetric polar diagrams as shown in Fig~\ref{fig:protocol}b.

\section*{Effect of the oscillations on the granular structure.}
\begin{figure*}
    \centering
    \includegraphics[width=0.9\linewidth]{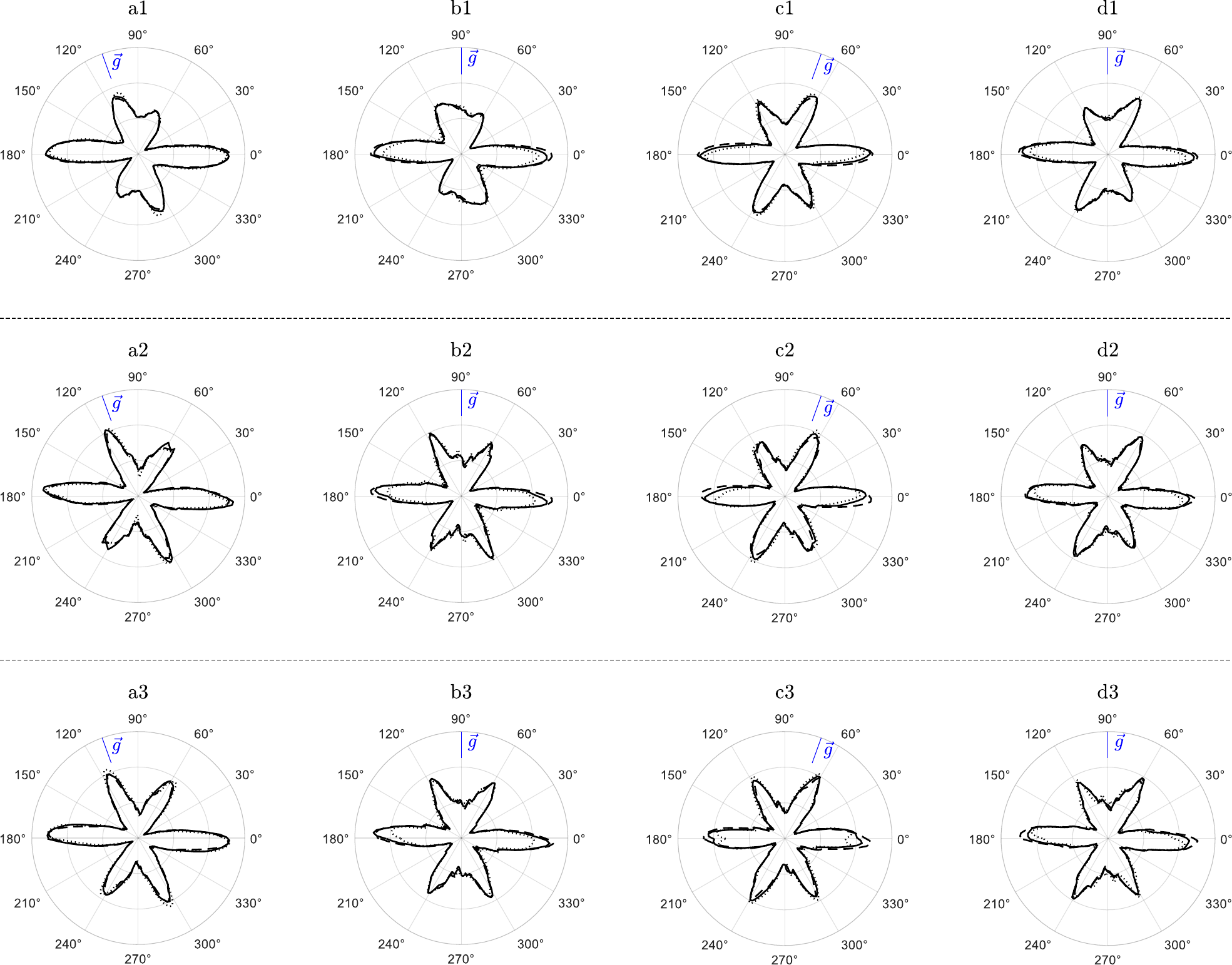}
    \caption{Evolution of the structure of the granular packing during the first three oscillations of the preparation protocol. The positions, $a_i$,$b_i$, $c_i$, $d_i$ are defined in Fig.\ref{fig:protocol}.
       }
    \label{fig:struct}
\end{figure*}

Figure \ref{fig:struct} details the evolution of neighboring-grain orientation over the first three oscillation cycles. The polar diagram $a_1$ corresponds to the granular structure immediately after the preparation avalanches. As explained in the main text, the lobes at $0$--$180^\circ$ reflects the developpement parallel grain layers during during the preparation avalanches, and the $120^\circ$ lobe is associated to the formation of force chains. Diagrams $b_1$ and $c_1$ show how this structure evolves as the packing is tilted in the direction opposite to that corresponding to $a_1$, with the development of a lobe at $60^\circ$ while the lobe at $120^\circ$ progressively disappears. This provides a clear indication of the reorientation of first-neighbor directions induced by the oscillatory motion. Note that the inclination range from $b_1$--$c_1$ corresponds to the loading path of the A0-R protocol, suggesting the structural changes at the origin of the large number of precursors observed for this protocol.

For the protocol A0-F, State $b_1$ corresponds to the stage at which the sample is about to be reloaded in the same direction as the preparation avalanches. Only minor structural changes are observed as a result of bringing the sample back to the horizontal position, which explains the absence of precursors: the existing structure (force chains) remains compatible with the applied loading direction.

The $d_1$--$a_2$ path corresponds to the loading of samples A1-F towards avalanche. As for A0-R, the numerous precursors are observed during the reorientation of the majority of first-neighbor directions, in this case from $60^\circ$ to $120^\circ$. Again, the structural changes accompanying precursor activity may arise from a reorientation of the force-chain network.

However, repeated oscillations progressively compact the sample, driving it towards a state in which the reorientation of first-neighbor directions becomes increasingly unlikely due to the tendency to form a more hexagonal network ecause of compaction. This progressive structural stabilization suppresses precursor activity, as observed along the $d_3$--$a_4$ path corresponding to the loading of samples A3-F towards $\theta_a$.

\end{document}